\RequirePackage{ifpdf}
\ifpdf 
\documentclass[pdftex]{sigma}
\else
\documentclass{sigma}
\fi

\newcommand{\pa}{\partial}
\newcommand{\la}{\lambda}
\newcommand{\tpa}{\tilde{\partial}}

\newcommand{\cA}{\mathcal{A}}
\newcommand{\cB}{\mathcal{B}}

\newcommand{\cF}{\mathcal{F}}

\newcommand{\cP}{\mathcal{P}}
\newcommand{\cS}{\mathcal{S}}

\newcommand{\bcP}{\boldsymbol{\mathcal{P}}}

\newcommand{\bA}{\boldsymbol{A}}
\newcommand{\bB}{\boldsymbol{B}}
\newcommand{\bC}{\boldsymbol{C}}

\newcommand{\bI}{\boldsymbol{I}}
\newcommand{\bJ}{\boldsymbol{J}}
\newcommand{\bK}{\boldsymbol{K}}

\newcommand{\bQ}{\boldsymbol{Q}}
\newcommand{\bR}{\boldsymbol{R}}
\newcommand{\bS}{\boldsymbol{S}}

\newcommand{\bU}{\boldsymbol{U}}
\newcommand{\bV}{\boldsymbol{V}}
\newcommand{\bW}{\boldsymbol{W}}
\newcommand{\bX}{\boldsymbol{X}}
\newcommand{\bXi}{\boldsymbol{\Xi}}
\newcommand{\bY}{\boldsymbol{Y}}
\newcommand{\bZ}{\boldsymbol{Z}}

\newcommand{\bbC}{\mathbb{C}}
\newcommand{\bbE}{\mathbb{E}}
\newcommand{\bbI}{\mathbb{I}}
\newcommand{\bbN}{\mathbb{N}}

\newcommand{\bbR}{\mathbb{R}}

\newcommand{\bbZ}{\mathbb{Z}}

\newcommand{\tXi}{\tilde{\Xi}}

\newcommand{\tbU}{\tilde{\boldsymbol{U}}}
\newcommand{\tbV}{\tilde{\boldsymbol{V}}}

\renewcommand{\d}{\mathrm{d}}
\newcommand{\bd}{\bar{\mathrm{d}}}

\newcommand{\imag}{\mathrm{i}}

\newcommand{\bfs}{\boldsymbol{s}}

\numberwithin{equation}{section}

\begin{document}

\allowdisplaybreaks

\renewcommand{\PaperNumber}{055}

\FirstPageHeading

\ShortArticleName{Bidif\/ferential Calculus Approach to AKNS Hierarchies and Their Solutions}

\ArticleName{Bidif\/ferential Calculus Approach\\ to AKNS Hierarchies and Their Solutions}

\Author{Aristophanes DIMAKIS~$^\dag$ and Folkert M\"ULLER-HOISSEN~$^\ddag$}

\AuthorNameForHeading{A. Dimakis and F. M\"uller-Hoissen}

\Address{$^\dag$~Department of Financial and Management Engineering,
 University of the Aegean,\\
 \hphantom{$^\dag$}~41, Kountourioti Str., GR-82100 Chios, Greece}
\EmailD{\href{mailto:dimakis@aegean.gr}{dimakis@aegean.gr}}

\Address{$^\ddag$~Max-Planck-Institute for Dynamics and Self-Organization, \\
 \hphantom{$^\ddag$}~Bunsenstrasse 10, D-37073 G\"ottingen, Germany}
\EmailD{\href{mailto:folkert.mueller-hoissen@ds.mpg.de}{folkert.mueller-hoissen@ds.mpg.de}}

\ArticleDates{Received April 12, 2010, in f\/inal form June 21, 2010;  Published online July 16, 2010}

\Abstract{We express AKNS hierarchies, admitting reductions to matrix NLS and matrix mKdV hierarchies,
in terms of a bidif\/ferential graded algebra. Application of a universal result in this framework
quickly generates an inf\/inite family of exact solutions, including e.g.\ the matrix solitons in the
focusing NLS case. Exploiting a general Miura transformation, we recover the generalized
Heisenberg magnet hierarchy and establish a corresponding solution formula for it.
Simply by exchanging the roles of the two derivations of the bidif\/ferential graded algebra,
we recover ``negative f\/lows'', leading to an extension of the respective hierar\-chy.
In this way we also meet a matrix and vector version of the short pulse equation and also
the sine-Gordon equation. For these equations corresponding solution formulas are also derived.
In all these cases the solutions are parametrized in terms of matrix data that have to
satisfy a certain Sylvester equation.}


\Classification{37J35; 37K10; 16E45}

\section{Introduction}

A unif\/ication of some integrability aspects and solution generating techniques has recently
been achieved for a wide class of ``integrable'' partial dif\/ferential or dif\/ference equations (PDEs)
in the framework of bidif\/ferential graded algebras \cite{DMH08bidiff}. The hurdle to take
is to f\/ind a bidif\/ferential calculus (i.e.\ bidif\/ferential graded algebra) associated with
the respective PDE. In particular, a~surprisingly simple result (Theorem 3.1 in
\cite{DMH08bidiff} and Theorem~\ref{theorem:main} below)
then generates a (typically large) class of exact solutions. This has been elaborated in detail for
matrix NLS systems in a~recent work \cite{DMH10NLS}. The present work extends some of these
results to a corresponding hierarchy and moreover to related hierarchies. It demonstrates
how to deal with whole hierarchies instead of only single equations or systems
in the bidif\/ferential calculus approach and shows moreover that certain relations between
hierarchies f\/ind a nice explanation in this framework.
Except for certain specializations, we deal with ``non-commutative equations'', i.e.\ we
treat the dependent variables as non-commutative matrices, and the solution formulas that
we present respect this fact.

In Section~\ref{sec:basic} we introduce some basic structures needed in the sequel.
Section~\ref{sec:AKNShier} presents a bidif\/ferential calculus for a matrix AKNS hierarchy.
We derive a class of solutions of the latter and address some reductions of the hierarchy.
In Sections~\ref{sec:rec_hier} and~\ref{sec:comb_hier} we show that, in the bidif\/ferential
calculus framework, a ``reciprocal'' \cite{Nijh88} or ``negative'' extension of the hierarchy
naturally appears. ``Negative f\/lows'' have been considered previously via negative powers of a~recursion operator (see e.g.~\cite{Fuch+Foka81,Vero91,Tracy+Widom96,JZZ09}
and also \cite{DGS98,Kamc+Pavl02,AFGZ00,AGZ06b} for other aspects). In our picture, these rather emerge
as ``mixed equations'', bridging between the ordinary hierarchy and a ``purely negative'' counterpart.

Section~\ref{sec:dualAKNS} elaborates this program for a ``dual hierarchy''. Here we recover
in the bidif\/ferential calculus framework in particular a well-known duality or gauge
equivalence between the (matrix) NLS and (generalized) Heisenberg magnet hierarchies \cite{Hasi72,Zakh+Thak79,Ishi82,Wada+Sogo83,Tsuc+Wada99WKI,Fadd+Takh87,vdLCN89,Gerd+GrahG04}.
Section~\ref{sec:concl} contains some concluding remarks.

\section{Basic structures}
\label{sec:basic}

\begin{definition}
A \emph{graded algebra} is an associative algebra $\Omega$ over $\bbC$ with a direct
sum decomposition
\begin{gather*}
     \Omega = \bigoplus_{r \geq 0} \Omega^r
\end{gather*}
into a subalgebra $\cA = \Omega^0$ and $\cA$-bimodules $\Omega^r$, such that
\begin{gather*}
     \Omega^r   \Omega^s \subseteq \Omega^{r+s}   .
\end{gather*}
\end{definition}

\begin{definition}
A \emph{bidifferential calculus} (or \emph{bidifferential graded algebra}) is a graded algebra
$\Omega$ equipped with two ($\bbC$-linear) graded derivations
$\d, \bar{\d} : \Omega \rightarrow \Omega$ of degree one (hence $\d \Omega^r \subseteq \Omega^{r+1}$,
$\bar{\d} \Omega^r \subseteq \Omega^{r+1}$), with the properties
\begin{gather*}
    \d \circ \d = 0  , \qquad
    \bar{\d} \circ \bar{\d} = 0   , \qquad
    \d \circ \bar{\d} + \bar{\d} \circ \d = 0   ,
\end{gather*}
and the graded Leibniz rule
\begin{gather*}
    \d(\chi   \chi') = (\d \chi)   \chi' + (-1)^r   \chi   \d \chi'    , \qquad
    \bar{\d}(\chi   \chi') = (\bar{\d}\chi)   \chi' + (-1)^r   \chi   \bar{\d} \chi'   ,
\end{gather*}
for all $\chi \in \Omega^r$ and $\chi' \in \Omega$.
\end{definition}

For any algebra $\cA$, a corresponding graded algebra is given by
\begin{gather}
    \Omega = \cA \otimes \bigwedge\big(\bbC^N\big)   ,    \label{Omega=AotimesC^2}
\end{gather}
where $\bigwedge(\bbC^N)$ denotes the exterior algebra of $\bbC^N$, $N>1$.
Def\/ining graded derivations~$\d$,~$\bar{\d}$ on~$\cA$, they extend
in an obvious way to $\Omega$ such that the Leibniz rule holds and elements of $\bigwedge(\bbC^N)$
are treated as constants with respect to $\d$ and $\bar{\d}$.

Given a bidif\/ferential calculus, it turns out that the equation
\begin{gather}
    \bar{\d} \d \phi = \d \phi \wedge \d \phi   ,  \qquad \mbox{where} \quad
     \phi \in \cA   ,    \label{phi_eq}
\end{gather}
has various integrability properties \cite{DMH08bidiff,DMH10NLS}. By choosing a suitable
bidif\/ferential calculus, this equation covers in particular the familiar selfdual
Yang--Mills equation (in one of its gauge-reduced potential versions), but also e.g.
discrete integrable equations~\cite{DMH08bidiff,DMH10NLS}. In the next section we
demonstrate that, by choosing an appropriate bidif\/ferential calculus,
(\ref{phi_eq}) reproduces matrix AKNS hierarchies.

The (modif\/ied) \emph{Miura transformation}
\begin{gather}
     [\bd g - (\d g)   \Delta ]   g^{-1} = \d \phi   ,  \label{Miura}
\end{gather}
where $\bd \Delta = (\d \Delta)   \Delta$, is a hetero-B\"acklund transformation between
(\ref{phi_eq}) and the \emph{dual equation}\footnote{Introducing
$B = [\bd g - (\d g)   \Delta ]   g^{-1}$, this equation reads $\d B =0$, and as
a consequence, taking also $\bd \Delta = (\d \Delta)   \Delta$ into account, we f\/ind
that $\bd B = B \wedge B$. If, as in familiar cases, this (partial)
zero curvature equation implies $B = (\bd g')   g'^{-1}$, then (\ref{g_eq}) is
gauge-equivalent to $\d [ (\bd g')   g'^{-1}] =0$, so that the term involving $\Delta$
in (\ref{g_eq}) can be generated by a gauge transformation. It is nevertheless helpful
to consider the modif\/ied equation (\ref{g_eq}) in order to accommodate more easily
certain examples of integrable equations in this formalism \cite{DMH08bidiff}.}
\begin{gather}
    \d \left( [ \bd g - (\d g)   \Delta ]  g^{-1} \right) = 0    . \label{g_eq}
\end{gather}
In the present work we concentrate on the case where $\Delta=0$.
We note that (\ref{Miura}) and (\ref{g_eq}) are equivalent if $\d$ has trivial cohomology.
But this is in general not the case.

Exchanging $\d$ and $\bd$ in (\ref{phi_eq}), we get a dif\/ferent equation.
Dealing with hierarchies, such an exchange leads to what we call the \emph{reciprocal hierarchy}.
If $\Delta=0$, exchanging $\d$ and $\bd$ in (\ref{g_eq}), simply amounts to replacing $g$ by $g^{-1}$.

\section{AKNS hierarchies}
\label{sec:AKNShier}

Let $\cB_0$ be the algebra of complex smooth functions of independent variables $t_1,t_2,t_3,\ldots$, $\cB$~an extension by certain operators (specif\/ied below),
and $\cA = \mathrm{Mat}(m,m,\cB)$, $m>1$. Let $m= m_1+m_2$ with $m_i \in \bbN$, and let
$\cP$ be the projection
\begin{gather}
    \cP = \cP_{(m_1,m_2)}
        = \left( \begin{array}{cc} I_{m_1} & 0_{m_1 \times m_2} \\
              0_{m_2 \times m_1} & 0_{m_2 \times m_2} \end{array} \right)   ,
          \label{projection_matrix}
\end{gather}
where $I_m$ denotes the $m \times m$ identity matrix. If the dimension is obvious from
the context, we will simply denote it by $I$.
It will also be convenient to introduce the matrix
\begin{gather*}
    J = J_{(m_1,m_2)} = 2  \cP - I_m    .
\end{gather*}

\subsection{NLS system}
A particular bidif\/ferential calculus on $\cA$ is determined by
\begin{gather}
     \d f = [\cP,f]   \zeta_1 + [\cP \pa_x,f]   \zeta_2   , \qquad
    \bd f = f_x  \zeta_1 + \frac{1}{2}   [\pa_{t_2} + \pa_x^2,f]   \zeta_2   ,
                     \label{NLS_bidiff}
\end{gather}
where $\zeta_1$, $\zeta_2$ is a basis of $\bigwedge^1(\bbC^2)$, and we set $x=t_1$.
Here $\cB$ is the extension of $\cB_0$ by the partial derivative operator $\pa_x$.
Evaluation of (\ref{phi_eq}) yields
\begin{gather}
    [\cP,\phi_{t_2}] = \{ \cP,\phi_{x x} \} + 2   (\cP \phi_x)   [\cP,\phi]
        - 2   [\cP,\phi]   \phi_x \cP   ,      \label{NLS_phi_eq}
\end{gather}
using the familiar notation for commutator and anti-commutator.
The block-decomposition
\begin{gather}
    \phi = J   \left(\begin{array}{cc} p & q \\ \bar{q} & \bar{p} \end{array}\right)
         = \left(\begin{array}{rr} p & q \\ -\bar{q} & -\bar{p} \end{array}\right)
            ,  \label{phi_decomp}
\end{gather}
(where $p$, $\bar{p}$, $q$, $\bar{q}$ have size $m_1 \times m_1$, $m_2 \times m_2$, $m_1 \times m_2$, $m_2 \times m_1$,
respectively), results in the NLS system\footnote{This matrix NLS system apparently
f\/irst appeared in~\cite{Zakh+Shab74}. See also the list of references in~\cite{DMH10NLS} and
in addition~\cite{Zakh80,Kono80,GGK05,Gerd+GrahG10}.}
\begin{gather}
    q_{t_2} = q_{xx} - 2 q   \bar{q}   q   , \qquad
    \bar{q}_{t_2} = - \bar{q}_{xx} + 2   \bar{q}  q   \bar{q}   ,
    \label{NLS_sys}
\end{gather}
together with
\begin{gather}
    p_x = - q   \bar{q}   , \label{p_x=-qr}
\end{gather}
where we set a ``constant'' of integration to zero. As a consequence of the form of~$\cP$,
there is no equation for $\bar{p}$. Though at this point we could simply set it to zero,
this would be inconsistent with further methods used in this work
(cf.\ Remark~\ref{rem:intermed}).

\subsection{Extension to a hierarchy}
\label{subsec:AKNShier}

Another bidif\/ferential calculus on $\cA$ is determined by
\begin{gather}
    \d f = [\cP \bbE_\la,f]   \zeta_1 + [\cP \bbE_\mu,f]   \zeta_2   , \qquad
    \bd f = \la^{-1} [\bbE_\la,f]   \zeta_1 + \mu^{-1} [\bbE_\mu,f]   \zeta_2   .
    \label{AKNS_bidiff_hier}
\end{gather}
Here $\bbE_\la$ and $\bbE_\mu$ are commuting invertible operators, which also commute
with $\cP$, and $\cB$ is the extension of $\cB_0$ by these operators.
Introducing $f_{[\la]} = \bbE_\la f   \bbE_\la^{-1}$ and
$f_{-[\la]} = \bbE_\la^{-1} f   \bbE_\la$, (\ref{phi_eq}) results in
\begin{gather}
 \left( \la^{-1} I - \cP\phi + \phi_{-[\la]} \cP \right)_{-[\mu]}
    \left( \mu^{-1} I - \cP\phi + \phi_{-[\mu]} \cP \right)
 \nonumber    \\
\qquad{}=  \left( \mu^{-1} I - \cP \phi + \phi_{-[\mu]} \cP \right)_{-[\la]}
     \left( \la^{-1} I - \cP \phi + \phi_{-[\la]} \cP \right)   .
              \label{AKNS_hier_funct}
\end{gather}
In the following, $\bbE_\la$ will be chosen as the \emph{Miwa shift} operator,
hence
\begin{gather*}
    f_{\pm [\la]}(t_1,t_2,t_3,\ldots) = f(t_1 \pm \la, t_2 \pm \la^2/2, t_3 \pm \la^3/3, \ldots)
    ,
\end{gather*}
with an arbitrary constant $\la$ (see e.g.~\cite{DMH06func}). The generating equation\footnote{In some
publications such an equation has been called a ``functional representation'' of the corresponding
hierarchy, see \cite{DMH06func} and the references cited therein. It seems to be more appropriate
to call it a ``generating (dif\/ferential) equation''.}
(\ref{AKNS_hier_funct}) already appeared in \cite{DMH06func} and we recall some
consequences from this reference.
Expanding (\ref{AKNS_hier_funct}) in powers of the arbitrary constants (indeterminates) $\la$ and $\mu$,
we recover (\ref{NLS_phi_eq}) as the coef\/f\/icient of $\la^1 \mu^0$.
Decomposing the matrix $\phi$ into blocks according to (\ref{phi_decomp}), this results in
\begin{gather}
  \left( \la^{-1} I - p + p_{-[\la]}\right)_{-[\mu]}
    \left( \mu^{-1} I -p +p_{-[\mu]}\right) + (q \bar{q})_{-[\mu]}
     \nonumber\\
\qquad{} = \left( \mu^{-1} I - p +p_{-[\mu]}\right)_{-[\la]}
    \left( \la^{-1} I -p +p_{-[\la]} \right) + (q \bar{q})_{-[\la]}   ,
          \label{preAKNShier_p}
\end{gather}
and
\begin{gather}
    \la^{-1} (q-q_{-[\la]}) + p_{-[\la]}   q
   =  \mu^{-1} (q-q_{-[\mu]}) + p_{-[\mu]}   q   , \label{preAKNShier_q} \\
    \la^{-1} (\bar{q}_{[\la]}-\bar{q}) + \bar{q}   p_{[\la]}
   =  \mu^{-1} (\bar{q}_{[\mu]} - \bar{q}) + \bar{q}   p_{[\mu]}   .
                \label{preAKNShier_bq}
\end{gather}
Again, there is no equation for $\bar{p}$.
Since the last equations separate with respect to $\la$ and~$\mu$, they imply
\begin{gather}
    \la^{-1} (q - q_{-[\la]}) - q_x - (p - p_{-[\la]})   q = 0   , \nonumber\\
    \la^{-1} (\bar{q} - \bar{q}_{-[\la]}) - \bar{q}_{-[\la],x}
    + \bar{q}_{-[\la]} (p - p_{-[\la]}) = 0   .    \label{q,bq_eqs}
\end{gather}
Multiplying the f\/irst equation from the right by $\bar{q}_{-[\la]}$, the second
from the left by $q$, using (\ref{p_x=-qr}) and adding the resulting equations, we obtain
\begin{gather}
    (p - p_{-[\la]} + \la   q   \bar{q}_{-[\la]})_x
  = [p - p_{-[\la]} + \la   q   \bar{q}_{-[\la]}, p - p_{-[\la]} ]   .
          \label{pre_p_eq}
\end{gather}
If we set integration constants to zero, this leads to
\begin{gather}
     \la^{-1} (p - p_{-[\la]}) = - q   \bar{q}_{-[\la]}   .    \label{p_eq}
\end{gather}
Using this equation, (\ref{q,bq_eqs}) becomes
\begin{gather}
    \la^{-1} (q - q_{-[\la]}) -q_x + \la   q   \bar{q}_{-[\la]}   q = 0   ,
    \qquad
     \la^{-1} (\bar{q}_{[\la]} - \bar{q}) - \bar{q}_x - \la   \bar{q}   q_{[\la]}   \bar{q} = 0
      ,    \label{AKNS_hier_func}
\end{gather}
which are generating equations for a hierarchy that contains the (matrix) NLS system.
Together with (\ref{p_x=-qr}), this leads to
\begin{gather}
 (\la^{-1} I -p + p_{-[\la]})_{-[\mu]}
    (\mu^{-1} I - p + p_{-[\mu]}) - (p_x)_{-[\mu]}
 \nonumber\\
\qquad{} =  (\mu^{-1} I - p + p_{-[\mu]})_{-[\la]}
    (\la^{-1} I - p + p_{-[\la]}) - (p_x)_{-[\la]}   ,  \label{KPhier}
\end{gather}
which is a generating equation for the potential KP hierarchy
\cite{Bogd+Kono98,DMH06func}\footnote{Its f\/irst member is the potential
KP equation $( 4   p_{t_3} - p_{xxx} - 6   (p_x)^2 )_x - p_{t_2t_2} + 6   [p_x , p_{t_2} ] =0$.}.
If we think of $p$ as determined via (\ref{p_x=-qr}) in terms of $q$ and $\bar{q}$, then
the last equation is a consequence of (\ref{AKNS_hier_func}). Furthermore, the two equations
(\ref{q,bq_eqs}) are the linear respectively adjoint linear system
of the KP hierarchy (cf.~\cite{Kono+Stra91}) in the form of generating equations.

Let us recall that
\begin{gather}
    \bbE_\la = \exp\left( \sum_{n \geq 1} \frac{1}{n}   \la^n   \pa_{t_n} \right)
                 = \sum_{n \geq 0} \la^n   \boldsymbol{s}_n(\tilde{\pa})
          \qquad \mbox{where} \quad
    \tilde{\pa} = \left(\pa_{t_1}, \frac{1}{2} \pa_{t_2}, \frac{1}{3} \pa_{t_3}, \ldots\right)   ,
       \label{E->Schur}
\end{gather}
and $\boldsymbol{s}_n$ are the elementary Schur polynomials.
Expanding (\ref{AKNS_hier_func}) in powers of $\la$ thus leads to
\begin{gather}
  \boldsymbol{s}_{n}(-\tilde{\pa})(q) - q   \boldsymbol{s}_{n-2}(-\tilde{\pa})(\bar{q})   q =0,\nonumber \\
  \boldsymbol{s}_{n}(\tilde{\pa})(\bar{q}) - \bar{q}   \boldsymbol{s}_{n-2}(\tilde{\pa})(q)   \bar{q} = 0,
  \qquad  n=2,3,\ldots   .   \label{AKNS_hier_Schur}
\end{gather}
For $n=2$ we recover (\ref{NLS_sys}). For $n=3$, and after elimination of $t_2$-derivatives
using (\ref{NLS_sys}), we obtain the system
\begin{gather}
    q_{t_3} = q_{xxx} - 3   (q_x   \bar{q}   q + q   \bar{q}   q_x)   , \qquad
    \bar{q}_{t_3} = \bar{q}_{xxx} - 3   (\bar{q}_x   q   \bar{q} + \bar{q}   q   \bar{q}_x)   ,
    \label{mKdV_sys}
\end{gather}
which admits reductions to matrix KdV and matrix mKdV equations (see also e.g.~ \cite{Zakh80,Kono80,Atho+Ford87KdV}).
In the same way, any pair of equations in (\ref{AKNS_hier_Schur}) can be expressed
in the form
\begin{gather}
    q_{t_{n}} = Q_n(q,\bar{q},q_x,\bar{q}_x,\ldots,q_{x^n}, \bar{q}_{x^n})   ,  \qquad
  \bar{q}_{t_{n}} = \bar{Q}_n(q,\bar{q},q_x,\bar{q}_x,\ldots,q_{x^n}, \bar{q}_{x^n})   ,
        \label{resolved_hier}
\end{gather}
by use of the equations for $q_{t_k}$, $\bar{q}_{t_k}$, with $k=2,\ldots,n-1$.
For $n=4$, we f\/ind
\begin{gather*}
   q_{t_4}  =  q_{xxxx} - 4   q   \bar{q}   q_{xx} - 4   q_{xx}   \bar{q}   q
     - 2   q   \bar{q}_{xx}   q - 6   q_x   \bar{q}   q_x - 2   q   \bar{q}_x   q_x
     - 2   q_x   \bar{q}_x   q + 6   q   \bar{q}   q   \bar{q}   q   , \\
 \bar{q}_{t_4}  =  - \bar{q}_{xxxx} + 4   \bar{q}   q   \bar{q}_{xx} + 4   \bar{q}_{xx}   q   \bar{q}
     + 2   \bar{q}   q_{xx}   \bar{q} + 6   \bar{q}_x   q   \bar{q}_x
     + 2   \bar{q}   q_x   \bar{q}_x
     + 2   \bar{q}_x   q_x   \bar{q} - 6   \bar{q}   q   \bar{q}   q   \bar{q}   .
\end{gather*}

\begin{remark}
Expansion of (\ref{AKNS_bidiff_hier}) leads to
\begin{gather*}
    \d f  =  \sum_{m \geq 0} \la^m   [ \cP \bfs_m(\tpa) , f ]   \zeta_1
           + \sum_{n \geq 0} \mu^n   [ \cP \bfs_n(\tpa) , f ]   \zeta_2   , \\
    \bd f  =  \sum_{m \geq 1} \la^{m-1}   [ \bfs_m(\tpa) , f ]   \zeta_1
           + \sum_{n \geq 1} \mu^{n-1}   [ \bfs_n(\tpa) , f ]   \zeta_2   .
\end{gather*}
Selecting the terms with the same powers of $\la$ and $\mu$, this suggests to def\/ine
\begin{gather*}
    \d_{(m,n)} f = [ \cP \bfs_m(\tpa) , f ]   \zeta_1 + [ \cP \bfs_n(\tpa) , f ]   \zeta_2
                       , \qquad
    \bd_{(m,n)} f = [ \bfs_{m+1}(\tpa) , f ]   \zeta_1 + [ \bfs_{n+1}(\tpa) , f ]   \zeta_2   .
\end{gather*}
For any choice of non-negative integers $m$, $n$, this determines a bidif\/ferential calculus.
With $m=0$ and $n=1$, we recover (\ref{NLS_bidiff}).
\end{remark}

\begin{remark}
\label{rem:Lax_hier}
(\ref{phi_eq}) is the integrability condition of the linear equation
\begin{gather}
    \bd \Psi = (\d\phi)   \Psi + \nu   \d \Psi     \label{bidiff_linsys}
\end{gather}
for an $m \times m$ matrix $\Psi$, where $\nu$ is a constant
(cf.~\cite{DMH08bidiff}). If $\Psi$ is invertible, this is (\ref{Miura}) with $\Delta = \nu   I$.
Evaluation for the above bidif\/ferential calculus leads to
\begin{gather*}
    \la^{-1}(\Psi - \Psi_{-[\la]}) = (\cP \phi-\phi_{-[\la]}\cP)\Psi
         + \nu   (\cP \Psi - \Psi_{-[\la]} \cP)   .
\end{gather*}
In terms of $\psi$ given by
\begin{gather*}
  \Psi = \psi   \exp\left( -\sum_{n \geq 1} (\nu \cP)^n   t_n \right)   ,
\end{gather*}
this takes the form
\begin{gather*}
      \la^{-1} (\psi - \psi_{-[\la]})
    = (\nu \cP + \cP \phi - \phi_{-[\la]} \cP)   \psi    ,
\end{gather*}
which is a generating equation for all Lax pairs of the hierarchy.
Expanding in powers of $\la$, the f\/irst two members of this family of linear equations are
\begin{gather*}
    \psi_x = L   \psi   , \qquad
    \psi_{t_2} = M   \psi   ,
\end{gather*}
where
\begin{gather*}
     L  =  \nu   \cP + [\cP , \phi]
       = \left( \begin{array}{cc} \nu & q \\ \bar{q} & 0 \end{array} \right)   , \\
     M  =  \nu^2   \cP + \nu   [\cP , \phi] + [\cP , \phi]^2 + \{ \cP , \phi_x\}
       = \left( \begin{array}{cc} \nu^2 + q \bar{q} + 2   p_x & \nu   q + q_x \\
                \nu   \bar{q} - \bar{q}_x & \bar{q} q \end{array} \right)   .
\end{gather*}
This constitutes a Lax pair for the NLS system~(\ref{NLS_sys}). In order
to obtain a more common Lax pair for the NLS system, we have to eliminate~$p_x$
via (\ref{p_x=-qr}) and add a
constant times the identity matrix to~$L$ and~$M$, together with a redef\/inition of the
``spectral parameter''. See also~\cite{DMH10NLS}.
\end{remark}

\subsection{A class of solutions}
\label{subsec:NLShier_sol}
So far we def\/ined a bidif\/ferential calculus on $\mathrm{Mat}(m,m,\cB)$. In the following we need
to extend it to a larger algebra. The space of all matrices over $\cB$, with size greater or
equal to that of $n_0 \times n_0$ matrices,
\begin{gather*}
    \mathrm{Mat}_{n_0}(\cB) = \bigoplus_{n',n \geq n_0} \mathrm{Mat}(n',n,\cB)   ,
\end{gather*}
attains the structure of a complex algebra $\cA$ with the usual matrix product extended trivially
by setting $A B=0$ whenever the sizes of $A$ and $B$ do not match. For the example introduced
in the preceding subsection, we set $n_0=2$.
Let $\Omega$ be the corresponding graded algebra (\ref{Omega=AotimesC^2}).
For each $n \geq 2$ and a split $n=n_1 + n_2$, we choose a projection matrix $\cP_{(n_1,n_2)}$
of the form~(\ref{projection_matrix}).
Then we can extend the bidif\/ferential calculus def\/ined in~(\ref{AKNS_bidiff_hier}) to $\cA$ by
simply def\/ining the commutators appearing there appropriately, e.g.\ for an $n \times m$ matrix
$f$ we set
\begin{gather*}
    [\cP \bbE_\la , f] = \cP_{(n_1,n_2)} \bbE_\la f - f   \cP_{(m_1,m_2)}   \bbE_\la   .
\end{gather*}
For the following general result, see \cite{DMH08bidiff,DMH10NLS}.

\begin{theorem}
\label{theorem:main}
Let $(\Omega, \d, \bar{\d})$ be a bidifferential calculus with
$\Omega = \cA \otimes \bigwedge(\bbC^N)$ and $\cA = \mathrm{Mat}_{n_0}(\cB)$,
for some $n_0 \in \bbN$.
For fixed $n \geq n_0$, let $\bX, \bY \in \mathrm{Mat}(n,n,\cB)$ be solutions of
the linear equations
\begin{gather*}
   \bar{\d} \bX = (\d \bX)   \bS    , \qquad
      \bar{\d} \bY = (\d \bY)   \bS   ,        
\end{gather*}
and
\begin{gather}
    \bR   \bX - \bX   \bS = -\bQ   \bY   , \qquad
    \bQ = \tbV   \tbU    ,       \label{RX+QY=XS}
\end{gather}
with $\d$- and $\bar{\d}$-constant matrices $\bS,\bR \in \mathrm{Mat}(n,n,\cB)$,
$\tbU \in \mathrm{Mat}(m,n,\cB)$, $\tbV \in \mathrm{Mat}(n,m,\cB)$.
If~$\bX$ is invertible, then
\begin{gather}
     \phi = \tbU \bY \bX^{-1} \tbV  \in \mathrm{Mat}(m,m,\cB)    \label{phi=UPhiV}
\end{gather}
satisfies
\begin{gather}
     \bd \phi = (\d \phi)   \phi + \d \theta   ,  \label{phi_interim_eq}
\end{gather}
with some $m \times m$ matrix $\theta$. By application of $\d$, this then implies
that $\phi$ solves \eqref{phi_eq}.
\end{theorem}

Now we apply this theorem to the bidif\/ferential calculus associated with the
hierarchy introduced in Section~\ref{sec:AKNShier}. We f\/ix $n_1$, $n_2$, and write $\bcP$
for $\cP_{(n_1,n_2)}$.
The linear equation $\bar{\d} \bX = (\d \bX)   \bS$ is then equivalent to
\begin{gather*}
    \la^{-1} (\bX - \bX_{-[\la]}) = (\bcP \bX - \bX_{-[\la]} \bcP)   \bS   .
\end{gather*}
$\d$ and $\bd$-constancy of $\bS$ means that $\bS$ is constant in the usual sense
(i.e.\ does not depend on the independent variables $t_1,t_2,\ldots$) and satisf\/ies
$[\bcP, \bS] =0$, which restricts it to a block-diagonal matrix, i.e.\
$\bS = \mbox{block-diag}(S_1,S_2)$.
Decomposing $\bX = \bX_d + \bX_o$ into a block-diagonal and an of\/f-block-diagonal part,
and using $[\bcP, \bX_d] =0$, $[\bcP, \bX_o] = \bJ   \bX_o = - \bX_o   \bJ$ with
\begin{gather*}
    \bJ = J_{(n_1,n_2)} = 2   \bcP - I_n   ,
\end{gather*}
we obtain
\begin{gather*}
    (\bX_d - \bX_{d,-[\la]})(\bI - \la   \bcP \bS) = 0   , \qquad
    \bX_o   (\bI - \la   \bar{\bcP}   \bS) = \bX_{o,-[\la]} (\bI - \la   \bcP   \bS)   ,
\end{gather*}
where $\bar{\bcP}$ denotes the projection $\boldsymbol{\bcP} - \bJ$.
The f\/irst equation implies that $\bX_d$ is constant. We write $\bX_d = \bA_d$.
Noting that $\bcP   \bS$ and $\bar{\bcP}   \bS$ commute, the solution of the second equation is
\begin{gather*}
    \bX_o = \bA_o   e^{\sum_{k \geq 1}(\bar{\bcP} \bS)^k t_k}   e^{-\sum_{l \geq 1}(\bcP \bS)^l t_l}   ,
\end{gather*}
with a constant of\/f-block-diagonal matrix $\bA_o$, hence
\begin{gather*}
    \bX = \bA_d + \bA_o   \bXi  \qquad \mbox{where} \quad
    \bXi = e^{-\xi(\bS) \bJ}   , \qquad  \xi(\bS) = \sum_{k \geq 1} \bS^k t_k    .
\end{gather*}
A corresponding expression holds for $\bY$,
\begin{gather*}
    \bY = \bB_d + \bB_o   \bXi   .
\end{gather*}
Now (\ref{RX+QY=XS}) splits into the two parts
\begin{gather*}
    \bR   \bA_d + \tbV   \tbU   \bB_d = \bA_d   \bS    , \qquad
    \bR   \bA_o + \tbV   \tbU   \bB_o = \bA_o   \bS   .
\end{gather*}
Assuming that $\bA_d$ is invertible, we can solve the f\/irst of these equations for $\bR$
and use the resulting formula to eliminate $\bR$ from the second. This results in
\begin{gather}
    \bS \bK - \bK \bS = \bV \bU   ,             \label{Sylvester}
\end{gather}
where
\begin{gather}
    \bK = -\bA_d^{-1} \bA_o   , \qquad
    \bU = \tbU   (\bB_o + \bB_d \bK)   , \qquad
    \bV = \bA_d^{-1} \tbV     .   \label{bK,bU,bV}
\end{gather}
Note that $\bK = \bK_o$ and $J \bU= -\bU \bJ$ (whereas $J \tbU = \tbU \bJ$).
Next we evaluate (\ref{phi=UPhiV}),
\begin{gather*}
   \phi = \tbU   \bY   \bX^{-1}   \tbV
         = \tbU   (\bB_d + \bB_o \bXi)(\bA_d + \bA_o \bXi)^{-1} \tbV
        = (\tbU   \bB_d + \tbU   \bB_o \bXi)(\bI - \bK \bXi)^{-1} \bV  \nonumber \\
\phantom{\phi}{} = (\tbU   \bB_d + (\bU -\tbU \bB_d \bK )   \bXi)(\bI - \bK \bXi)^{-1} \bV
        = \bU   \bXi   (\bI - \bK   \bXi)^{-1} \bV + \tbU   \bB_d   \bV   .
\end{gather*}
Using the identity $(\bI - \bK \bXi)^{-1} = (\bI + \bK \bXi)(\bI - (\bK \bXi)^2)^{-1}$,
this decomposes into
\begin{gather}
    \phi_d = \phi_{d,0} + \bU \bXi   \bK \bXi   (\bI - (\bK \bXi)^2)^{-1}   \bV   ,
             \qquad \phi_{d,0} = \tbU   \bB_d   \bV   ,  \label{phi_d}
\end{gather}
and
\begin{gather}
    \phi_o = \bU   \bXi   (\bI - (\bK \bXi)^2)^{-1}   \bV   .  \label{phi_o}
\end{gather}
All this leads to the following result, which generalizes Proposition~5.1 in \cite{DMH10NLS}.

\begin{proposition}
\label{prop:hier_sol}
Let
\begin{center}
\begin{tabular}{c|c|c|c|c|c|c}
   \tsep{1pt}\bsep{1pt}   & $S$ & $\bar{S}$ & $U$ & $\bar{U}$ & $V$ & $\bar{V}$  \\
\hline
\tsep{1pt}\bsep{1pt}  size   & $n_1 \times n_1$ & $n_2 \times n_2$ & $m_1 \times n_2$ & $m_2 \times n_1$
        & $n_1 \times m_1$ & $n_2 \times m_2$
\end{tabular}
\end{center}
be constant complex matrices, and let $K$ $($of size $n_1 \times n_2)$ and
$\bar{K}$ $($of size $n_2 \times n_1)$ be solutions of the Sylvester equations
\begin{gather}
    S K + K \bar{S} = V U   , \qquad
    \bar{S} \bar{K} + \bar{K} S = \bar{V} \bar{U}   .
    \label{decomp_Sylvester}
\end{gather}
Then
\begin{gather}
   q = U   \bar{\Xi}   (I_{n_2} - \bar{K}   \Xi   K   \bar{\Xi})^{-1}   \bar{V}   ,
    \qquad
   \bar{q} = \bar{U}   \Xi   (I_{n_1} - K   \bar{\Xi}   \bar{K}   \Xi)^{-1}   V   ,
    \label{q,bq_sol}
\end{gather}
where
\begin{gather*}
    \Xi = e^{-\xi(S)}   , \qquad \bar{\Xi} = e^{\xi(-\bar{S})}   ,  \qquad
    \xi(S) = \sum_{k \geq 1} S^k t_k   ,
\end{gather*}
solve the hierarchy \eqref{AKNS_hier_func}. Furthermore,
\begin{gather}
    p = U   \bar{\Xi}   \bar{K}   \Xi   (I_{n_1} - K   \bar{\Xi}   \bar{K}   \Xi )^{-1}   V
        \label{p_sol}
\end{gather}
solves \eqref{p_eq} and then also the potential KP hierarchy \eqref{KPhier}.
\end{proposition}

\begin{proof} The expressions (\ref{decomp_Sylvester}), (\ref{q,bq_sol}) and (\ref{p_sol})
follow, respectively, from (\ref{Sylvester}), (\ref{phi_o}) and (\ref{phi_d}), by writing
\begin{gather}
    \bK = \left(\!\!\begin{array}{cc} 0 & K \\ \bar{K} & 0 \end{array}\!\!\right)   , \!\qquad
    \bS = \left(\!\!\begin{array}{cc} S & 0 \\ 0 & -\bar{S} \end{array}\!\!\right)   ,\! \qquad
    \bU = \left(\!\!\begin{array}{cc} 0 & U \\ -\bar{U} & 0 \end{array}\!\!\right)   , \!\qquad
    \bV = \left(\!\!\begin{array}{cc} V & 0 \\ 0 & \bar{V} \end{array}\!\!\right)    ,\!\!\!
             \label{bK,bS,bU,bV_matrixform}
\end{gather}
$\bXi = \mbox{block-diag}(\Xi,\bar{\Xi})$, and using (\ref{phi_decomp}). From
Theorem~\ref{theorem:main} we know that (\ref{q,bq_sol}) and (\ref{p_sol}) solve the
hierarchy equations (\ref{preAKNShier_p}), (\ref{preAKNShier_q}) and (\ref{preAKNShier_bq}).
It should be noticed, however, that on the way to the hierarchy (\ref{AKNS_hier_func})
the step to (\ref{p_eq}) involved a restriction. Hence we have to verify that $p$ given by
(\ref{p_sol}) actually solves (\ref{p_eq}).
Noting that
\begin{gather*}
     \Xi_{-[\la]} = \Xi   (I - \la   S)^{-1}   , \qquad
     \bar{\Xi}_{-[\la]} = \bar{\Xi}   (I + \la   \bar{S})   , \qquad
     Z   \Xi   K = K   \Bar{\Xi}   \bar{Z}   ,
\end{gather*}
where $I$ stands for the respective identity matrix, $Z = \Xi^{-1} - K   \bar{\Xi}   \bar{K}$
and $\bar{Z} = \bar{\Xi}^{-1} - \bar{K}   \Xi   K$, we have
\begin{gather*}
     Z_{-[\la]}
  =  \Xi^{-1} (I-\la   S) - K   \bar{\Xi}   (I + \la   \bar{S})   \bar{K}
  = Z - \la   ( \Xi^{-1}   S + K   \bar{\Xi}   \bar{S}   \bar{K} ) \\
 \phantom{Z_{-[\la]}}{}  =  Z - \la   ( \Xi^{-1}   S + K   \bar{\Xi}   [ \bar{V}   \bar{U} - \bar{K}   S ] )
  = Z   ( I-\la   S - \la   Z^{-1}   K   \bar{\Xi}   \bar{V}   \bar{U} ) \\
\phantom{Z_{-[\la]}}{} =  Z   ( I-\la   S - \la   \Xi   K   \bar{Z}^{-1}   \bar{V}   \bar{U} )   ,
\end{gather*}
where we used the second of equations (\ref{decomp_Sylvester}). Now we f\/ind that
$p = U   \bar{\Xi}   \bar{K}   Z^{-1}   V$ satisf\/ies
\begin{gather*}
     p - p_{-[\la]}
  =  U   \bar{\Xi}   \Big( \bar{K}   Z^{-1} Z_{-[\la]} - (I + \la   \bar{S})   \bar{K}
     \Big)   Z^{-1}_{-[\la]}   V \\
\phantom{p - p_{-[\la]}}{}
  =  U   \bar{\Xi}   \Big( \bar{K}   ( I -\la   S - \la   \Xi   K   \bar{Z}^{-1}   \bar{V}   \bar{U} )
     - (I + \la   \bar{S})   \bar{K} \Big)   Z^{-1}_{-[\la]}   V  \\
\phantom{p - p_{-[\la]}}{}
  =  - \la   U   \bar{\Xi}   \Big( \bar{K}   \Xi   K   \bar{Z}^{-1}   \bar{V}   \bar{U}
     + \underbrace{\bar{S}   \bar{K} + \bar{K}   S}_{=\bar{V}   \bar{U}} \Big)
         Z^{-1}_{-[\la]}   V  \\
\phantom{p - p_{-[\la]}}{}
  =  - \la   U   \bar{\Xi}   \big( \bar{K}   \Xi   K
     + \bar{Z} \big)   \bar{Z}^{-1}   \bar{V}   \bar{U}   Z^{-1}_{-[\la]}   V
  = - \la   q   \bar{q}_{-[\la]}   .\tag*{\qed}
\end{gather*}\renewcommand{\qed}{}
\end{proof}

There are matrix data for which the Sylvester equations (\ref{decomp_Sylvester}) have no solution.
But if $S$ and $-\bar{S}$ have no common eigenvalue, they admit a solution, irrespective of the
right hand side, and this solution is then unique (see e.g. Theorem~4.4.6 in \cite{Horn+John91}).

\begin{remark}
\label{rem:intermed}
Evaluated for the bidif\/ferential calculus (\ref{AKNS_bidiff_hier}),
(\ref{phi_interim_eq}) reads
\begin{gather*}
     \la^{-1} ( \phi - \phi_{-[\la]} ) = (\cP \phi - \phi_{-[\la]} \cP)   \phi
            + \cP \theta - \theta_{-[\la]} \cP   .
\end{gather*}
Using (\ref{phi_decomp}) and a corresponding block-decomposition for $\theta$,
\begin{gather*}
     \theta = \left( \begin{array}{rr} s & r \\ -\bar{r} & - \bar{s} \end{array} \right)   ,
\end{gather*}
this equation splits into the system
\begin{gather*}
    \la^{-1} (p - p_{-[\la]})  =  (p - p_{-[\la]})   p - q \bar{q} + s - s_{-[\la]}   , \\
    \la^{-1} (q - q_{-[\la]})  =  (p - p_{-[\la]})   q - q \bar{p} + r   , \\
    \la^{-1} (\bar{q} - \bar{q}_{-[\la]})  =  - \bar{q}_{-[\la]} p - \bar{r}_{-[\la]}   , \\
    \la^{-1} (\bar{p} - \bar{p}_{-[\la]})  =  - \bar{q}_{-[\la]} q    ,
\end{gather*}
which implies $q_x = - q \bar{p} + r$, $\bar{q}_x = - \bar{q} p - \bar{r}$, and
$p_x = -q \bar{q}$. With the help of these equations we recover (\ref{q,bq_eqs}) and
thus also (\ref{pre_p_eq}).
We observe that the solution generating method based on Theorem~\ref{theorem:main}
also imposes dif\/ferential equations on $\bar{p}$. The corresponding generating equation
is actually the direct counterpart of (\ref{p_eq}), which the solutions determined by
Proposition~\ref{prop:hier_sol} satisfy.
\end{remark}

\begin{remark}
\label{rem:cohomology}
If $\theta$ in (\ref{phi_interim_eq}) is not restricted and if we have \emph{trivial}
$\d$-cohomology, then (\ref{phi_interim_eq}) is equivalent to (\ref{phi_eq}).
But $\d$ given by (\ref{AKNS_bidiff_hier}) has \emph{non-trivial} cohomology.
Indeed, the following 1-form is $\d$-closed but not $\d$-exact,
\begin{gather*}
  \rho = \bbE_\la \left( \begin{array}{cc} a(\la) & 0 \\ 0 & \star \end{array} \right)   \zeta_1
       + \bbE_\mu \left( \begin{array}{cc} b(\mu) & 0 \\ 0 & \star \end{array} \right)   \zeta_2
           .
\end{gather*}
Here $a$, $b$ only depend on the parameter $\la$, respectively~$\mu$, and a star stands
for an arbitrary entry (not restricted in the dependence on the independent variables
and the respective parameter). Adding $\rho$ on the r.h.s.\ of (\ref{phi_interim_eq}) would
achieve equivalence with~(\ref{phi_eq}).
\end{remark}

\subsection{Reductions}
\label{subsec:red}
Let us consider the substitution $\pa_{t_{2k}} \mapsto -\pa_{t_{2k}}$, $k=1,2,\ldots$,
in the hierarchy equations (\ref{AKNS_hier_Schur}). By inspection of (\ref{E->Schur}), it
implies $\boldsymbol{s}_n(\tilde{\pa}) \mapsto (-1)^n   \boldsymbol{s}_n(-\tilde{\pa})$,
hence maps (\ref{AKNS_hier_Schur}) into
\begin{gather*}
  \boldsymbol{s}_{n}(\tilde{\pa})(q) - q   \boldsymbol{s}_{n-2}(\tilde{\pa})(\bar{q})   q = 0,\nonumber \\
  \boldsymbol{s}_{n}(-\tilde{\pa})(\bar{q}) - \bar{q}   \boldsymbol{s}_{n-2}(-\tilde{\pa})(q)   \bar{q} = 0,
  \qquad  n=2,3,\ldots   ,   
\end{gather*}
which has the same ef\/fect as
\begin{gather*}
   q \mapsto \epsilon   \bar{q}^\omega   , \qquad  \bar{q} \mapsto \epsilon   q^\omega   ,
\end{gather*}
where $\epsilon = \pm 1$, and $\omega$ is the involution  on the algebra of matrices either given by the identity map
or by complex conjugation ($\omega = \ast$), or the anti-involution either given by transposition
($\omega=\intercal$) or by Hermitian conjugation ($\omega=\dagger$) (see also \cite{DMH10NLS})\footnote{Note
that this may require restricting the size of the matrices. For example, if $\omega$ is complex conjugation,
we need $m_1 = m_2$.}.

It follows that the \emph{odd-time} part of the hierarchy, expressed in the form (\ref{resolved_hier}),
is consistent with the reduction condition
\begin{gather}
     \bar{q} = \epsilon   q^\omega   ,  \label{red}
\end{gather}
which reduces any of its pairs to a single member. In particular, (\ref{mKdV_sys}) becomes the
matrix mKdV equation
\begin{gather*}
    q_{t_3} = q_{xxx} - 3   \epsilon   (q_x   q^\omega   q + q   q^\omega   q_x),
    \qquad   \epsilon = \pm 1   .    
\end{gather*}
The reduced hierarchy is therefore a matrix mKdV hierarchy.

After the replacement
\begin{gather*}
    t_{2k} \mapsto \imag   t_{2k}, \qquad  k=1,2,\ldots   ,
\end{gather*}
with $\imag = \sqrt{-1}$, so that $\pa_{t_{2k}} \mapsto - \imag   \pa_{t_{2k}}$, also the
\emph{even-time} equations of the hierarchy are consistent with the above reduction (\ref{red}),
provided we choose for $\omega$ complex conjugation ($\ast$) or Hermitian conjugation ($\dagger$),
so that $\imag^\omega = - \imag$. Then (\ref{NLS_sys}) becomes the matrix NLS equation
\begin{gather*}
    \imag   q_{t_2} = - q_{xx} + 2   \epsilon   q   q^\omega   q,
    \qquad   \epsilon = \pm 1   .     
\end{gather*}
This matrix version of the NLS equation apparently f\/irst appeared in \cite{Zakh+Shab74}.
The corresponding reduced hierarchy is a matrix NLS hierarchy.

Since Proposition~\ref{prop:hier_sol} provides a class of solutions of the original
hierarchy in terms of matrix data, we should address the question what kind of
constraints a reduction imposes on the latter.

\subsubsection{Reduction using an involution}

If $\omega$ is one of the involutions specif\/ied above, setting
\begin{gather}
    \bar{S} = S^\omega   , \qquad
    \bar{U} = \epsilon   \epsilon'   U^\omega   , \qquad
    \bar{V} = \epsilon'   V^\omega   , \qquad
    \bar{K} = \epsilon   K^\omega   ,    \label{invol_conds_matrix-data}
\end{gather}
with $\epsilon' = \pm 1$, and arranging that $\bar{\Xi} = \Xi^\omega$,
i.e.\ $\xi(-S^\omega) = \xi(-\bar{S}) = - \xi(S)^\omega$,
achieves the reduction condition (\ref{red}). This forces us to set
\begin{gather*}
    m_1 = m_2   , \qquad  n_1 = n_2   .
\end{gather*}
Renaming $m_i$ to $m$ and $n_i$ to $n$, this leads to the following
consequence of Proposition~\ref{prop:hier_sol}.

\begin{proposition}
Let $S,U,V$ be constant $n \times n$, $m \times n$, respectively $n \times m$ matrices,
and let $K$ be a solution of the Sylvester equation
\begin{gather*}
    S   K + K   S^\omega = V   U   .
\end{gather*}
Then
\begin{gather*}
 q = \pm U   \Xi^\omega   (I_n - \epsilon   K^\omega   \Xi   K   \Xi^\omega)^{-1}   V^\omega
       , \qquad \mbox{where} \quad
 \Xi = e^{-\xi(S)}
\end{gather*}
with
\begin{gather*}
    \xi(S) = \left\{ \begin{array}{ll}
    \displaystyle \sum_{k \geq 0} S^{2k+1}   t_{2k+1}\qquad &
    \mbox{if} \quad \mbox{$\omega=\mathrm{id}$ \quad $($``real'' mKdV$)$}, \vspace{1mm}\\
     \displaystyle        \sum_{k \geq 0} S^{2k+1}   t_{2k+1} + \imag   \sum_{k \geq 1} S^{2k}   t_{2k}
\qquad &  \mbox{if} \quad
                 \mbox{$\omega=\mathrm{\ast}  $ \quad $($NLS-mKdV$)$},
                     \end{array}
                     \right.
\end{gather*}
solves the $m \times m$ matrix ``real'' mKdV, respectively NLS-mKdV hierarchy.
\end{proposition}

If the involution is given by complex conjugation, in the focusing NLS case the solutions
obtained from this proposition include matrix (multiple) solitons. See \cite{DMH10NLS} for
corresponding results for the respective NLS equation, the f\/irst member of the
hierarchy\footnote{In \cite{DMH10NLS} we used a bidif\/ferential calculus dif\/ferent from the
one chosen in the present work. But the resulting expressions for exact solutions are
the same.}.

\subsubsection{Reduction using an anti-involution}
In this case the reduction condition (\ref{red}) can be implemented on the solutions
determined by Proposition~\ref{prop:hier_sol} by setting
\begin{gather*}
    \bar{S} = S^\omega   , \qquad
          U = V^\omega   , \qquad
    \bar{U} = \epsilon   \bar{V}^\omega   , \qquad
    K^\omega = K   , \qquad
    \bar{K}^\omega = \bar{K}   ,
\end{gather*}
and arranging again that $\bar{\Xi} = \Xi^\omega$. For the anti-involutions specif\/ied above,
we are forced to set $n_1 = n_2$, which we rename to $n$. Then we have the following result.

\begin{proposition}
Let $S$, $V$, $\bar{V}$ be constant matrices of size $n \times n$, $n \times m_1$ and $n \times m_2$,
respectively. Let $K$, $\bar{K}$ be $($with respect to $\omega)$ Hermitian solutions of the
Sylvester equations\footnote{Here the previous matrix $\bar{K}$ has been redef\/ined with a factor $\epsilon$.}
\begin{gather*}
     S   K + K   S^\omega = V   V^\omega   , \qquad
     S^\omega   \bar{K} + \bar{K}   S = \bar{V}   \bar{V}^\omega   .
\end{gather*}
Then
\begin{gather*}
  q = V^\omega   \Xi^\omega   (I_n - \epsilon   \bar{K}   \Xi   K   \Xi^\omega)^{-1}   \bar{V}
        , \qquad \mbox{where} \quad
  \Xi = e^{-\xi(S)}
\end{gather*}
with
\begin{gather*}
    \xi(S) = \left\{ \begin{array}{ll}
    \displaystyle \sum_{k \geq 0} S^{2k+1}   t_{2k+1} \qquad &
    \mbox{if} \quad \mbox{$\omega=\intercal$ \quad $($``real'' mKdV$)$},\vspace{1mm}
     \\
          \displaystyle     \sum_{k \geq 0} S^{2k+1}   t_{2k+1} + \imag   \sum_{k \geq 1} S^{2k}   t_{2k}
                    \qquad & \mbox{if} \quad
                                \mbox{$\omega=\dagger$ \quad $($NLS-mKdV$)$},
                     \end{array} \right.
\end{gather*}
solves the $m_1 \times m_2$ matrix ``real'' mKdV, respectively NLS-mKdV hierarchy.
\end{proposition}

If the involution is given by Hermitian conjugation, in the focusing NLS case the solutions
obtained from the last proposition include matrix (multiple) solitons. Corresponding results
for the respective NLS equation, the f\/irst member of the hierarchy, have been obtained
in~\cite{DMH10NLS}.

\section{The reciprocal AKNS hierarchy}
\label{sec:rec_hier}
Exchanging the roles of $\d$ and $\bd$ in (\ref{AKNS_bidiff_hier}), we have
\begin{gather}
    \d f = \la^{-1} [\bar{\bbE}_\la,f]   \zeta_1 + \mu^{-1} [\bar{\bbE}_\mu,f]   \zeta_2
             , \qquad
    \bd f = [\cP \bar{\bbE}_\la,f]   \zeta_1 + [\cP \bar{\bbE}_\mu,f]   \zeta_2   .
    \label{reciprocal_NLS_bidiff_hier}
\end{gather}
Here the Miwa shift operator is def\/ined in terms of a new set of independent variables,
$\bar{t}_i$, $i=1,2,\ldots$, and $f \in \mathrm{Mat}(m,m,\bar{\cB})$, where $\bar{\cB}$ is
the algebra of smooth functions of these variables, extended by the Miwa shifts.
Now (\ref{phi_eq}) results in
\begin{gather}
    \Big( \cP - \la^{-1} (\phi - \phi_{-\bar{[\la]}} )_{-\bar{[\mu]}} \Big)
   \Big( \cP - \mu^{-1} (\phi - \phi_{-\bar{[\mu]}} ) \Big) \nonumber \\
 \qquad {}=  \Big( \cP - \mu^{-1} (\phi - \phi_{-\bar{[\mu]}} )_{-\bar{[\la]}} \Big)
    \Big( \cP - \la^{-1} (\phi - \phi_{-\bar{[\la]}} ) \Big)   .
                 \label{rec_NLS_hier_funct}
\end{gather}
In terms of
\begin{gather*}
   \varphi = \phi - \bar{x}   \cP   ,  
\end{gather*}
where $\bar{x} = \bar{t}_1$, the above equation reduces to
\begin{gather}
    (\varphi - \varphi_{-\bar{[\la]}} )_{-\bar{[\mu]}}   (\varphi - \varphi_{-\bar{[\mu]}} )
  = (\varphi - \varphi_{-\bar{[\mu]}} )_{-\bar{[\la]}}   (\varphi - \varphi_{-\bar{[\la]}} )  .
       \label{KPhier_nonlin_part}
\end{gather}
Expanding this in powers of $\la$ and $\mu$, to order $\la^2 \mu$ it yields
\begin{gather}
   ( \varphi_{\bar{x}}{}^2 )_{\bar{x}} - [ \varphi_{\bar{x}} , \varphi_{\bar{t}_2} ] = 0   .
          \label{KP_nlin_part}
\end{gather}
This is nothing but the nonlinear part of the potential KP equation.
More generally, (\ref{KPhier_nonlin_part})~is the nonlinear part of the potential KP hierarchy
as obtained from~(\ref{KPhier})\footnote{The nonlinear part can be extracted via the scaling limit
$\epsilon \to 0$ of the potential KP hierarchy, assuming $p(t_1,t_2,\ldots) = \varphi(\bar{t}_1,\bar{t}_2,\ldots)$ for the KP variable $p$, where $\bar{t}_n = \epsilon^n   t_n$.}.
To order $\la^3 \mu$, (\ref{KPhier_nonlin_part})~yields
\begin{gather*}
     [\varphi_{\bar{x}} , \varphi_{\bar{t}_3} ]
   = \frac{3}{4}   \left( \{ \varphi_{\bar{x}} , \varphi_{\bar{t}_2} \}
      + \frac{1}{3}   [ \varphi_{\bar{x}} , \varphi_{\bar{x}\bar{x}} ] \right)_{\bar{x}}   .
\end{gather*}

\begin{remark}
\label{rem:KPnlin-HM}
(\ref{KP_nlin_part}) is related to the generalized Heisenberg magnet (gHM) equation
$2   \cS_{\bar{t}_2} = [\cS , \cS_{\bar{x}\bar{x}}]$
(see e.g.~\cite{Zakh+Thak79,Cher96,Golu+Soko99,Gerd+GrahG04}) for an $m \times m$ matrix $\cS$ as follows.
The gHM equation is compatible with the constraint $\cS^2 = I$.
Writing $\cS = 2   \varphi_{\bar{x}} - I$, the constraint reads $\varphi_{\bar{x}}{}^2 = \varphi_{\bar{x}}$,
which implies $\varphi_{\bar{x}}   \varphi_{\bar{x}\bar{x}}   \varphi_{\bar{x}} = 0$,
and the gHM equation becomes
$\varphi_{\bar{t}_2} = [\varphi_{\bar{x}} , \varphi_{\bar{x}\bar{x}}]$, setting a constant of
integration to zero. Then $[ \varphi_{\bar{x}} , \varphi_{\bar{t}_2} ] = (\varphi_{\bar{x}}{}^2)_{\bar{x}}$,
hence $\varphi$ satisf\/ies the nonlinear part of the potential KP equation.
\end{remark}

\begin{remark}
\label{rem:Lax_recipr_hier}
The general linear equation (\ref{bidiff_linsys})\footnote{Using the def\/initions
of Section~\ref{subsec:AKNShier}, this linear equation reads
$\d \Psi = (\bd \phi)   \Psi + \nu   \bd \Psi$, since in the present section we
exchanged the def\/initions of $\d$ and $\bd$ relative to those of Section~\ref{subsec:AKNShier}.}
leads to
\begin{gather*}
    \cP \Psi - \Psi_{-\bar{[\la]}} \cP = \la^{-1} \big(\phi-\phi_{-\bar{[\la]}}\big)   \Psi
         + \frac{\nu}{\la}   \big(\Psi - \Psi_{-\bar{[\la]}}\big)   .
\end{gather*}
Setting
\begin{gather*}
  \Psi = \psi   \exp\left( -\sum_{n \geq 1} (\nu^{-1} \cP)^n   \bar{t}_n \right)   ,
\end{gather*}
this takes the form
\begin{gather*}
      \psi - \psi_{-\bar{[\la]}}
    = - \nu^{-1}   \big(\varphi - \varphi_{-\bar{[\la]}}\big)   \psi   ,
\end{gather*}
which is a generating equation for all Lax pairs of the reciprocal hierarchy.
The f\/irst two members of this family of linear equations are
\begin{gather*}
    \psi_{\bar{x}} = \bar{L}   \psi   , \qquad
    \psi_{\bar{t}_2} = \bar{M}   \psi   ,
\end{gather*}
where
\begin{gather*}
   \bar{L}  =  - \nu^{-1}   \varphi_{\bar{x}}
       = \nu^{-1}   \left( \begin{array}{cc} 1 - p_{\bar{x}} & - q_{\bar{x}} \\
                \bar{q}_{\bar{x}} & \bar{p}_{\bar{x}} \end{array} \right)   , \\
   \bar{M}  =  \nu^{-2}   \varphi_{\bar{x}}{}^2 - \nu^{-1}   \varphi_{\bar{t}_2} \\
      \phantom{\bar{M}}{}
        =  \left( \begin{array}{cc} \nu^{-2}   (1+2   p_{\bar{x}}
            + p_{\bar{x}}{}^2 - q_{\bar{x}} \bar{q}_{\bar{x}} ) -\nu^{-1}   p_{\bar{t}_2} &
          \nu^{-2}   (q_{\bar{x}} + p_{\bar{x}} q_{\bar{x}}
             - q_{\bar{x}} \bar{p}_{\bar{x}} ) -\nu^{-1}   q_{\bar{t}_2} \vspace{1mm}\\
         - \nu^{-2}   (\bar{q}_{\bar{x}} + \bar{q}_{\bar{x}} p_{\bar{x}}
             - \bar{p}_{\bar{x}} \bar{q}_{\bar{x}} ) + \nu^{-1}   \bar{q}_{\bar{t}_2} &
               \nu^{-2}   (\bar{p}_{\bar{x}}{}^2
               - \bar{q}_{\bar{x}} q_{\bar{x}} ) + \nu^{-1}   \bar{p}_{\bar{t}_2} \end{array} \right)   .
\end{gather*}
\end{remark}

\subsection{A class of solutions}
We apply again Theorem~\ref{theorem:main}. Using (\ref{reciprocal_NLS_bidiff_hier}),
$\bar{\d} \bX = (\d \bX)   \bS$ takes the form
\begin{gather*}
     \la^{-1} \big( \bX - \bX_{-\bar{[\la]}} \big) = \big( \cP \bX - \bX_{-\bar{[\la]}} \cP \big)   \bS^{-1}   ,
\end{gather*}
assuming that $\bS$ is invertible. Decomposition as in Section~\ref{subsec:NLShier_sol} leads to
\begin{gather*}
    \bX = \bA_d + \bA_o   \bXi   , \qquad   \bY = \bB_d + \bB_o   \bXi
     \qquad \mbox{where} \quad
     \bXi = e^{-\xi(\bS)}   , \qquad \xi(\bS) = \sum_{k \geq 1} \bS^{-k}   \bar{t}_k   .
\end{gather*}
These are the same formulas we obtained in Section~\ref{subsec:NLShier_sol}, but with $\bS$
replaced by $\bS^{-1}$. From (\ref{RX+QY=XS}) we obtain, however, the same Sylvester equation,
$\bS   \bK - \bK   \bS = \bV   \bU$, using the same def\/initions as in (\ref{bK,bU,bV}).
Furthermore, we obtain again (\ref{phi_d}) and (\ref{phi_o}), and thus the following
counterpart of Proposition~\ref{prop:hier_sol}.

\begin{proposition}
\label{prop:recipr_hier_sol}
Let
\begin{center}
\begin{tabular}{c|c|c|c|c|c|c}
    \tsep{1pt}\bsep{1pt}  & $S$ & $\bar{S}$ & $U$ & $\bar{U}$ & $V$ & $\bar{V}$  \\
\hline
 \tsep{1pt}\bsep{1pt} size   & $n_1 \times n_1$ & $n_2 \times n_2$ & $m_1 \times n_2$ & $m_2 \times n_1$
        & $n_1 \times m_1$ & $n_2 \times m_2$
\end{tabular}
\end{center}
be constant complex matrices, where $S$ and $\bar{S}$ are invertible, and let $K$
$($of size $n_1 \times n_2)$ and $\bar{K}$ $($of size $n_2 \times n_1)$ be solutions of
the Sylvester equations
\begin{gather*}
    S K + K \bar{S} = V U   , \qquad
    \bar{S} \bar{K} + \bar{K} S = \bar{V} \bar{U}   .
\end{gather*}
Then $\varphi = \phi - \bar{x}   \cP$, where $\phi$ is given by \eqref{phi_decomp}
with the components
\begin{alignat*}{3}
 &  q = U   \bar{\Xi}   (I_{n_2} - \bar{K}   \Xi   K   \bar{\Xi})^{-1}   \bar{V}   ,
    \qquad &&
   \bar{q} = \bar{U}   \Xi   (I_{n_1} - K   \bar{\Xi}   \bar{K}   \Xi)^{-1}   V   , & \nonumber \\
&    p = U   \bar{\Xi}   \bar{K}   \Xi   (I_{n_1} - K   \bar{\Xi}   \bar{K}   \Xi )^{-1}   V   ,
     \qquad &&
   \bar{p} = \bar{U}   \Xi   K   \bar{\Xi}   (I_{n_2} - \bar{K}   \Xi   K   \bar{\Xi})^{-1}
               \bar{V}   , &
\end{alignat*}
and
\begin{gather*}
    \Xi = e^{-\xi(S)}   , \qquad \bar{\Xi} = e^{\xi(-\bar{S})}   ,  \qquad
    \xi(S) = \sum_{k \geq 1} S^{-k} \bar{t}_k   ,
\end{gather*}
solves the hierarchy \eqref{KPhier_nonlin_part}.
\end{proposition}

\section{The combined hierarchy}
\label{sec:comb_hier}
The bidif\/ferential calculus determined by
\begin{gather}
 \d f  =  [\cP \bbE_{\la_1},f]   \zeta_1 + [\cP \bbE_{\la_2},f]   \zeta_2
           + \mu_1^{-1} [\bar{\bbE}_{\mu_1} , f]   \bar{\zeta}_1
           + \mu_2^{-1} [\bar{\bbE}_{\mu_2} , f]   \bar{\zeta}_2
              , \nonumber \\
 \bd f  =  \la_1^{-1} [\bbE_{\la_1},f]   \zeta_1 + \la_2^{-1} [\bbE_{\la_2} , f]   \zeta_2
           + [\cP \bar{\bbE}_{\mu_1} , f]   \bar{\zeta}_1
           + [\cP \bar{\bbE}_{\mu_2} , f]   \bar{\zeta}_2   ,
                      \label{NLS_comb_bidiff}
\end{gather}
contains (\ref{AKNS_bidiff_hier}) and (\ref{reciprocal_NLS_bidiff_hier}), and thus
combines the corresponding hierarchies. Here $\zeta_1$, $\zeta_2$, $\bar{\zeta}_1$, $\bar{\zeta}_2$
is a basis of $\bigwedge^1(\bbC^4)$, $\la_i$ and $\mu_i$ are indeterminates,
and $f \in \mathrm{Mat}(m,m,\cB)$, where
$\cB$ is now the algebra of complex smooth functions of independent variables
$t_1,t_2,t_3,\ldots$, $\bar{t}_1,\bar{t}_2, \ldots$, extended by the Miwa shift operators.
Then (\ref{phi_eq}) is equivalent to (\ref{AKNS_hier_funct}), (\ref{rec_NLS_hier_funct}),
and\footnote{This equation alone is obtained
from (\ref{phi_eq}) using the bidif\/ferential calculus determined by
$\d f = [\cP \bbE_\la , f ]   \zeta + \mu^{-1}   [\bar{\bbE}_\mu , f ]   \bar{\zeta}$ and
$\bd f = \la^{-1}   [ \bbE_\la , f ]   \zeta + [\cP \bar{\bbE}_\mu , f ]   \bar{\zeta}$.
The remaining equations of the hierarchy can be recovered from the linear system with
this (restricted) bidif\/ferential calculus.}
\begin{gather*}
   \Big( \cP - \mu^{-1} (\phi - \phi_{-\bar{[\mu]}})_{-[\la]} \Big)
         \Big( \la^{-1} I - (\cP \phi - \phi_{-[\la]} \cP) \Big)    \nonumber \\
 \qquad{}  =  \Big( \la^{-1} I - (\cP \phi - \phi_{-[\la]} \cP)_{-\bar{[\mu]}} \Big)
        \Big( \cP - \mu^{-1} (\phi - \phi_{-\bar{[\mu]}}) \Big)   .
\end{gather*}
To order $\la^0$, respectively $\mu^0$, this yields
\begin{gather}
    \la^{-1} (\phi - \phi_{-[\la]})_{\bar{x}}
   =  (\cP - \phi_{-[\la],\bar{x}}) (\cP \phi - \phi_{-[\la]} \cP)
    - (\cP \phi - \phi_{-[\la]} \cP) (\cP - \phi_{\bar{x}})   , \label{comb1} \\
    \mu^{-1} (\phi - \phi_{-\bar{[\mu]}})_x
 =  (\cP - \mu^{-1} (\phi - \phi_{\bar{[\mu]}}) )   [\cP, \phi]
      - [\cP , \phi ]_{-\bar{[\mu]}}   (\cP - \mu^{-1} (\phi - \phi_{-\bar{[\mu]}}) )   .
                     \label{comb2}
\end{gather}
To order $\la^0 \mu^0$ we have
\begin{gather*}
    \phi_{x \bar{x}} = [   \cP - \phi_{\bar{x}} , [ \cP, \phi ]   ]   .
\end{gather*}
Using (\ref{phi_decomp}), this becomes
\begin{gather}
     q_{x \bar{x}} + p_{\bar{x}}   q + q   \bar{p}_{\bar{x}} = q   , \qquad
     \bar{q}_{x \bar{x}} + \bar{p}_{\bar{x}}   \bar{q} + \bar{q}   p_{\bar{x}} = \bar{q}   ,
           \label{mixed1a}
\end{gather}
and $p_{x \bar{x}} =- (q   \bar{q})_{\bar{x}}$, $\bar{p}_{x \bar{x}} =- (\bar{q}   q)_{\bar{x}}$,
which integrates to
\begin{gather}
    p_x = - q   \bar{q}   , \qquad
  \bar{p}_x = - \bar{q}   q   ,    \label{mixed1b}
\end{gather}
setting constants of integration to zero.

To order $\la$, (\ref{comb1}) leads to
\begin{gather*}
     \varphi_{t_2 \bar{x}} = - \left[ \varphi_{\bar{x}} , \{ \cP , \varphi_x \} + [\cP , \varphi]^2 \right]
          ,
\end{gather*}
which decomposes into
\begin{gather*}
  p_{t_2 \bar{x}}  =  - [p_{\bar{x}} , q \bar{q} ] - q_{\bar{x}} \bar{q}_x + q_x \bar{q}_x   , \\
  \bar{p}_{t_2 \bar{x}}  =  [\bar{p}_{\bar{x}} , \bar{q} q ] + \bar{q}_{\bar{x}} q_x
                           - \bar{q}_x q_{\bar{x}}   , \\
  q_{t_2 \bar{x}}  =  (p_{\bar{x}} -1)   q_x + q_x   \bar{p}_{\bar{x}}
         + q_{\bar{x}} \bar{q} q + q \bar{q} q_{\bar{x}}   , \\
   \bar{q}_{t_2 \bar{x}}  =  - \bar{q}_x   (p_{\bar{x}} -1) - q_x   \bar{p}_{\bar{x}}
         - \bar{q}_{\bar{x}} q \bar{q} - \bar{q} q \bar{q}_{\bar{x}}   .
\end{gather*}
Furthermore, to order $\mu$ (\ref{comb2}) yields
\begin{gather*}
     \varphi_{\bar{t}_2 x} = - [ \varphi_{\bar{t}_2} , [ \cP , \varphi ] ]
              - \{ \varphi_{\bar{x}} , [\cP , \varphi_{\bar{x}} ] \}    ,
\end{gather*}
which leads to
\begin{gather*}
     q_{\bar{t}_2 x} = - p_{\bar{t}_2} q - q \bar{p}_{\bar{t}_2} - p_{\bar{x}} q_{\bar{x}}
             + q_{\bar{x}} \bar{p}_{\bar{x}}   , \qquad
     \bar{q}_{\bar{t}_2 x} = - \bar{p}_{\bar{t}_2} \bar{q} - \bar{q} p_{\bar{t}_2}
             - \bar{p}_{\bar{x}} \bar{q}_{\bar{x}}
             + \bar{q}_{\bar{x}} p_{\bar{x}}   .    
\end{gather*}

In the same way as in the preceding sections, we arrive at the following result.

\begin{proposition}
\label{prop:comb_hier_sol}
Let
\begin{center}
\begin{tabular}{c|c|c|c|c|c|c}
  \tsep{1pt}\bsep{1pt}    & $S$ & $\bar{S}$ & $U$ & $\bar{U}$ & $V$ & $\bar{V}$  \\
\hline
 \tsep{1pt}\bsep{1pt} size   & $n_1 \times n_1$ & $n_2 \times n_2$ & $m_1 \times n_2$ & $m_2 \times n_1$
        & $n_1 \times m_1$ & $n_2 \times m_2$
\end{tabular}
\end{center}
be constant complex matrices, where $S$ and $\bar{S}$ are invertible, and let $K$
$($of size $n_1 \times n_2)$ and $\bar{K}$ $($of size $n_2 \times n_1)$ be solutions of
the Sylvester equations
\begin{gather*}
    S K + K \bar{S} = V U   , \qquad
    \bar{S} \bar{K} + \bar{K} S = \bar{V} \bar{U}   .
\end{gather*}
Then $\phi$ given by \eqref{phi_decomp} with the components
\begin{alignat*}{3}
&    q = U   \bar{\Xi}   (I_{n_2} - \bar{K}   \Xi   K   \bar{\Xi})^{-1}   \bar{V}  ,
    \qquad &&
   \bar{q} = \bar{U}   \Xi   (I_{n_1} - K   \bar{\Xi}   \bar{K}   \Xi)^{-1}   V   , & \nonumber \\
 &  p = U   \bar{\Xi}   \bar{K}   \Xi   (I_{n_1} - K   \bar{\Xi}   \bar{K}   \Xi )^{-1}   V  ,
     \qquad &&
   \bar{p} = \bar{U}   \Xi   K   \bar{\Xi}   (I_{n_2} - \bar{K}   \Xi   K   \bar{\Xi})^{-1}
               \bar{V}   , &
\end{alignat*}
where
\begin{gather*}
    \Xi = e^{-\xi(S)}   , \qquad \bar{\Xi} = e^{\xi(-\bar{S})}   ,  \qquad
    \xi(S) = \sum_{k \geq 1} S^k t_k + \sum_{k \geq 1} S^{-k} \bar{t}_k   ,
\end{gather*}
solves all equations of the combined hierarchy.
\end{proposition}

This simply extends Propositions~\ref{prop:hier_sol} and~\ref{prop:recipr_hier_sol}
by adding the respective expressions for~$\xi(S)$.

\subsection{A reduction}
Let $q$, $\bar{q}$, $p$, $\bar{p}$ be square matrices of the same size. Setting
\begin{gather}
     \bar{q} = \epsilon   q   , \qquad
     \bar{p} = p   , \qquad \mbox{where} \quad \epsilon = \pm 1   ,
              \label{comb_hier_red1}
\end{gather}
the system (\ref{mixed1a}), (\ref{mixed1b}) reduces to
\begin{gather*}
 p_x = - \epsilon   q^2   , \qquad
 q_{x \bar{x}} = \frac{1}{2} (I - 2 p_{\bar{x}} )   q + \frac{1}{2} q   (I - 2 p_{\bar{x}})   .
\end{gather*}
This reduction corresponds to the choice $\omega = \mathrm{id}$ in Section~\ref{subsec:red} and
extends more generally to the ``odd f\/lows'' of the combined hierarchy.
Imposing the conditions~(\ref{invol_conds_matrix-data}) on the matrix data of
Proposition~\ref{prop:comb_hier_sol}, then leads to the following result.

\begin{proposition}
\label{prop:red1_odd_comb_hier}
Let $S$, $U$, $V$ be constant $n \times n$, $m \times n$, respectively $n \times m$ matrices,
and let $K$ be a solution of the Sylvester equation $S   K + K   S = V   U$.
Then
\begin{gather*}
    q = \pm U   \Xi   \big(I_n - \epsilon   (K   \Xi)^2 \big)^{-1}   V
          , \qquad
    p = \epsilon   U   \Xi   K   \Xi
          \big(I_n - \epsilon   (K   \Xi)^2 \big)^{-1}   V   ,
\end{gather*}
where
\begin{gather*}
  \Xi = e^{-\xi(S)}   , \qquad
  \xi(S) = \sum_{k \geq 1} S^{2k-1} t_{2k-1} + \sum_{k \geq 1} S^{-2k+1} \bar{t}_{2k-1}
              ,
\end{gather*}
solve the odd part of the combined hierarchy with the reduction condition~\eqref{comb_hier_red1}.
\end{proposition}

\subsubsection{Short pulse equation}
Let us impose the additional condition that $p$ is a scalar times the identity matrix. Then we have
\begin{gather*}
    p_x = - \epsilon   q^2   , \qquad
    q_{x \bar{x}} = (I - 2 p_{\bar{x}} )   q   .
\end{gather*}
Writing
\begin{gather}
     p = \frac{1}{2} (\bar{x} - z)   I_m   ,   \label{p-z}
\end{gather}
with a new dependent scalar variable $z$, the last system is turned into
\begin{gather}
     z_x   I = 2   \epsilon   q^2   , \qquad
     q_{x\bar{x}} = z_{\bar{x}}   q   .    \label{z_x,q_xbx}
\end{gather}
In terms of $u(x,z)$ given by
\begin{gather}
       u(x,z(x,\bar{x})) = 2   q(x,\bar{x})   ,    \label{u=2q}
\end{gather}
we obtain
\begin{gather*}
   2   q_x = u_x + z_x   u_z = u_x + \frac{\epsilon}{2}   u^2   u_z
           , \qquad
   z_{\bar{x}}   u = 2   q_{x \bar{x}}
                    = z_{\bar{x}}   \left( u_x + \frac{\epsilon}{2}   u^2   u_z \right)_z   .
\end{gather*}
The change of independent variables requires $z_{\bar{x}} \neq 0$.
The last equation is then equivalent to
\begin{gather}
     u_{x z} = u - \frac{\epsilon}{2}   \big(u^2   u_z\big)_z   ,   \label{SIT}
\end{gather}
which is a matrix version of the \emph{short pulse equation}. The latter apparently f\/irst appeared
in~\cite{Rabelo89} (see also~\cite{BRT89,Sako+Sako07}) and was later derived as an approximation for
the propagation of ultra-short pulses in nonlinear media~\cite{Schae+Wayn04}.
It was further studied in particular in \cite{Sako+Sako05,Sako+Sako06,Brunelli06,KBK07a,KBK08,Parkes08,PKB08,Matsuno09}.
Of course, we have to take the additional condition into account that $u^2$ has to be a
scalar times the identity matrix. This is achieved if
\begin{gather*}
     u = \sum_{i=1}^r u_i   \boldsymbol{e}_i   ,
\end{gather*}
where $\boldsymbol{e}_i   \boldsymbol{e}_j + \boldsymbol{e}_j   \boldsymbol{e}_i = 2 \eta_{ij}   I$
with $\eta_{ij} = \pm   \delta_{ij}$ (Clif\/ford algebra), since then $u^2 = \langle \boldsymbol{u} ,
\boldsymbol{u} \rangle   I$, where $\boldsymbol{u} = (u_1,\ldots,u_r)^\intercal$ and
$\langle \boldsymbol{u} , \boldsymbol{u} \rangle = \sum_{i=1}^r \eta_{ij}   u_i u_j$.
In this case (\ref{SIT}) becomes
\begin{gather}
    \boldsymbol{u}_{x z} = \boldsymbol{u}
   - \frac{\epsilon}{2}   (\langle \boldsymbol{u} , \boldsymbol{u} \rangle
                 \boldsymbol{u}_z)_z    .    \label{vSIT}
\end{gather}
This vector version of the short pulse equation is dif\/ferent from those considered
in~\cite{PKB08,Sakovich08}. In the following example we obtain an inf\/inite set of
exact solutions of the 2-component system via Proposition~\ref{prop:red1_odd_comb_hier}.

\begin{example}
\label{ex:preSPE_sol}
We can alternatively express the solution given in Proposition~\ref{prop:red1_odd_comb_hier}
in the a priori more redundant form
\begin{gather*}
    q = \pm U   \Xi   \big(I_n - \epsilon   (K   \tXi)^2 \big)^{-1}   V   , \qquad
    p = \epsilon   U   \tXi   K   \tXi
          \big(I_n - \epsilon   (K   \tXi)^2 \big)^{-1}   V   ,
\end{gather*}
with
\begin{gather*}
    S K + K S = V U   , \qquad  \tXi = C   \Xi   ,
\end{gather*}
where $C$ is any constant $n \times n$ matrix that commutes with~$S$. More precisely, the introduction
of~$C$ allows us to f\/ix some of the freedom in the choice of the coef\/f\/icients of the
matrices~$U$,~$V$. The following choices involve further restrictions, however.
We consider the case $m=2$, $n=2N$, and choose
\begin{gather*}
     S = \mathrm{diag}(s_1 I_2, \ldots, s_N I_2)   , \qquad
     U = \left( \begin{array}{ccc} \sigma_1 & \ldots & \sigma_1 \end{array} \right)   , \qquad
     V = \left( \begin{array}{c} I_2 \\ \vdots \\ I_2 \end{array} \right)   ,
\end{gather*}
where $\sigma_1$ is the respective Pauli matrix. The Sylvester equation is then solved by
\begin{gather*}
      K_{ij} = \frac{1}{s_i + s_j}   \sigma_1,  \qquad   i,j=1,\ldots,N   .
\end{gather*}
Choosing $C$ block-diagonal where the $2 \times 2$ blocks on the diagonal are a linear
combination of~$I_2$ and the Pauli matrix $\sigma_3 = \mathrm{diag}(1,-1)$, then $C$ commutes with~$S$.
Furthermore, it follows that $K \tXi$ consists of $2 \times 2$ blocks $(K \tXi)_{ij}$
which are linear combinations of the of\/f-diagonal Pauli matrices $\sigma_1$ and $\sigma_2$.
It further follows that $((K \tXi)^2)_{ij}$ is diagonal. As a consequence, the inverse of
$I_{2N} - \epsilon   (K   \tXi)^2$ also consists of diagonal $2 \times 2$
blocks\footnote{This is quite obvious if we regard the matrices as $N \times N$ matrices
over the commutative algebra of diagonal $2 \times 2$ matrices.}.
Because $U$ consists of of\/f-diagonal $2 \times 2$ blocks, we conclude that $q$ given by
the above formula is an of\/f-diagonal $2 \times 2$ matrix. Hence its square is proportional
to the identity matrix $I_2$. Since $N \in \bbN$ is arbitrary, we thus have an
inf\/inite family of exact solutions of the system~(\ref{vSIT}) with $r=2$, where the components
of the vector $\boldsymbol{u} = (u_1,u_2)^\intercal$ are given by $u = u_1 \sigma_1 + u_2 \sigma_2$.
Fig.~\ref{fig:vSPE_2soliton} shows a plot of a 2-soliton solution.

\begin{figure}[t]
\centerline{\includegraphics[width=16cm]{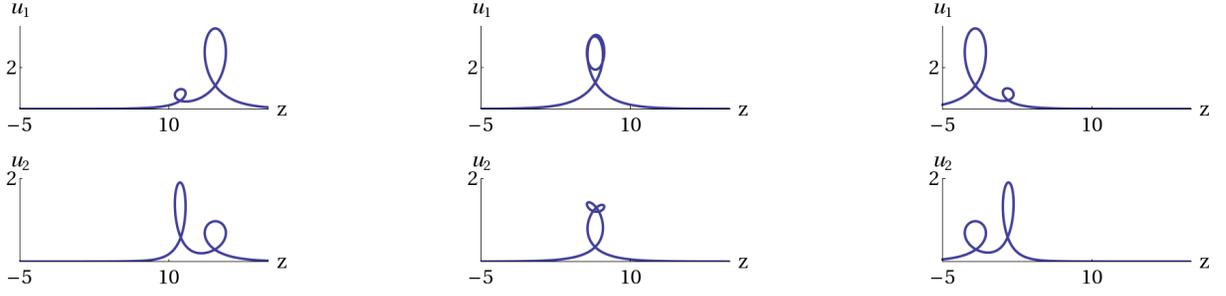}}
\caption{Parametric plots (with parameter $\bar{x}$) of the two components $u_1$ and $u_2$ for a
2-soliton solution of the 2-component short pulse equation (\ref{vSIT}) with $\epsilon=-1$,
determined by the data of Example~\ref{ex:preSPE_sol} with $N=2$, $s_1=1$, $s_2=2$ and
$C = \mathrm{diag}(1+3 \imag,1-3\imag, 2+\imag/2,2-\imag/2)$ at $x=-3,-1,1$.}\label{fig:vSPE_2soliton}
\end{figure}
\end{example}

\begin{remark}
In order to obtain a Lax pair for the short pulse equation, we start with
\begin{gather*}
     \psi_x = \left(L - \frac{\nu}{2} I\right)   \psi   , \qquad
     \psi_{\bar{x}} = \left(\bar{L} - \frac{1}{2 \nu} I\right)   \psi   ,
\end{gather*}
with $L$ and $\bar{L}$ taken from Remarks~\ref{rem:Lax_hier} and~\ref{rem:Lax_recipr_hier}, respectively. Without imposing a reduction, the integrability
conditions are~(\ref{mixed1a}) and~(\ref{mixed1b}) (modulo an integration with respect
to $\bar{x}$). Writing $\psi(x,\bar{x}) = \chi(x,z(x,\bar{x}))$, we f\/ind
\begin{gather*}
    \chi_z  =  (\bar{L} - \frac{1}{2 \nu} I)   \chi
           = \frac{1}{2 \nu} \left( \begin{array}{cc} I & -u_z \\ \epsilon   u_z & - I \end{array} \right)   \chi
               , \\
    \chi_x  =  (L - \frac{\nu}{2} I)   \chi - z_x   \chi_z
            = \frac{1}{2} \left( \begin{array}{cc} \nu & u \\
                  \epsilon   u & - \nu \end{array} \right) \chi - \frac{\epsilon}{2} u^2   \chi_z  \\
\phantom{\chi_x}{}
            =  \left( \begin{array}{cc} \frac{\nu}{2} I - \frac{\epsilon}{4\nu} u^2 &
                  \frac{1}{2} u + \frac{\epsilon}{4\nu} u^2   u_z \vspace{1mm}\\
                  \frac{\epsilon}{2} u - \frac{1}{4\nu} u^2   u_z &
                  -\frac{\nu}{2} I + \frac{\epsilon}{4\nu} u^2  \end{array} \right) \chi    ,
\end{gather*}
where we applied the reduction conditions~(\ref{comb_hier_red1}) and used (\ref{p-z}), (\ref{z_x,q_xbx})
and (\ref{u=2q}). Using the symmetry $u \mapsto -u$ of the short pulse equation, we recover the Lax pair
given in~\cite{Sako+Sako05} for the scalar case and with~$\epsilon =-1$.
\end{remark}

\section{Dual AKNS hierarchies}
\label{sec:dualAKNS}
 For the bidif\/ferential calculus determined by (\ref{AKNS_bidiff_hier}), the
Miura transformation (\ref{Miura}) with $\Delta=0$ takes the form
\begin{gather}
    \la^{-1} \big(g - g_{-[\la]}\big)   g^{-1} = \cP \phi - \phi_{-[\la]} \cP   ,
       \label{Miura_funct}
\end{gather}
and the \emph{dual equation} (\ref{g_eq}) becomes
\begin{gather*}
    \mu^{-1} \left( \cP   g_{-[\mu]}   g^{-1} - (g_{-[\mu]}   g^{-1})_{-[\la]}   \cP \right)
  = \la^{-1} \left( \cP   g_{-[\la]}   g^{-1} - (g_{-[\la]}   g^{-1})_{-[\mu]}   \cP \right)   .
\end{gather*}
Multiplying from the left by $g^{-1}_{-[\la]-[\mu]}$ and from the right by $g$, this
can be written as
\begin{gather}
    \mu^{-1} \left( g^{-1}_{-[\la]}   \cP   g - (g^{-1}_{-[\la]}   \cP   g )_{-[\mu]} \right)
  = \la^{-1} \left( g^{-1}_{-[\mu]}   \cP   g - (g^{-1}_{-[\mu]}   \cP   g)_{-[\la]} \right)   .
          \label{dualhier_funct1}
\end{gather}
To order $\mu^0$ this yields
\begin{gather}
    \la^{-1} \left( g^{-1} \cP   g - (g^{-1} \cP   g)_{-[\la]} \right)
  = ( g^{-1}_{-[\la]}   \cP   g )_x
      .     \label{dualhier_funct2}
\end{gather}
Applying a Miwa shift with $-[\mu]$ and subtracting the result from this equation, we obtain
\begin{gather*}
    \mu^{-1} \left( g^{-1}_{-[\la]}   \cP   g - (g^{-1}_{-[\la]}   \cP   g)_{-[\mu]} \right)_x\\
    \qquad
   =  \la^{-1}  \mu^{-1} \Big( g^{-1} \cP   g - (g^{-1} \cP   g)_{-[\la]}
   - (g^{-1} \cP   g)_{-[\mu]} + (g^{-1} \cP   g)_{-[\la]-[\mu]}  \Big)   .
\end{gather*}
Since the r.h.s.\ is symmetric in $\la$, $\mu$, this implies (\ref{dualhier_funct1}) up
to an $x$-independent ``constant of integration''. Hence~(\ref{dualhier_funct1})
reduces to~(\ref{dualhier_funct2}).
The f\/irst non-trivial equation resulting from an expansion of~(\ref{dualhier_funct2}) in powers of the indeterminate $\la$ is obtained as the term
linear in $\la$,
\begin{gather}
   (g^{-1} \cP   g)_{t_2} - (g^{-1} \cP   g)_{xx} = - 2   \left( (g^{-1})_x   \cP   g \right)_x
       .    \label{dualhier_t2eq}
\end{gather}

\subsection{Miura transformation and relation\\ with the generalized Heisenberg magnet model}
\label{subsec:gHM}
Introducing
\begin{gather*}
     \cS = g^{-1} J g
\end{gather*}
(see also \cite{vdLCN89}), we have the identities
\begin{gather}
    \cS_{t_n} = [\cS , g^{-1} g_{t_n} ],  \qquad   n=1,2,\ldots   , \label{cS_tn}
\end{gather}
and
\begin{gather}
    \cS^2 = I \quad \Rightarrow \quad
    \cS_x   \cS = - \cS   \cS_x \quad \Rightarrow \quad
    \cS_x{}^2 = - \frac{1}{2}   \{ \cS , \cS_{xx} \} \quad \Rightarrow \quad
    (\cS \cS_x)_x = \frac{1}{2}   [ \cS , \cS_{xx} ]   .
            \label{dh_cS-ids}
\end{gather}
The dual hierarchy equation (\ref{dualhier_t2eq}) can now be expressed as follows,
\begin{gather*}
    \cS_{t_2} = \cS_{xx} + 2   \left( g^{-1} g_x   (I + \cS) \right)_x   .
\end{gather*}
In order to express this solely in terms of $\cS$ we use the Miura transformation
(\ref{Miura_funct}), which to order $\la^0$ reads
\begin{gather}
     g_x   g^{-1}  = [\cP , \phi ] = \frac{1}{2}   [ J, \phi ]   .
          \label{dh_g-cP}
\end{gather}
This imposes the following condition on $g$,
\begin{gather}
     \big\{ J , g_x   g^{-1} \big\} = 0   ,    \label{J_anticom_g_xg^-1=0}
\end{gather}
which can be expressed as $\{ \cS , g^{-1} g_x \} = 0$, hence together with (\ref{cS_tn})
we have
\begin{gather}
    g^{-1} g_x = - \frac{1}{2}   \cS_x   \cS =: \cF^{(1)}   .  \label{dh_g^-1g_x->cS}
\end{gather}
Inserting this in the expression for $\cS_{t_2}$, and using the above identities,
leads to
\begin{gather}
    \cS_{t_2} = \cS_{xx} - \left( \cS_x   \cS   (I + \cS) \right)_x
              = \frac{1}{2}   [ \cS , \cS_{xx} ]
              = 2   \cF^{(1)}_x   ,  \label{dh_HMeq}
\end{gather}
which is the (generalized) Heisenberg magnet equation (see also Remark~\ref{rem:KPnlin-HM}).

\begin{remark}
The ordinary Heisenberg magnet equation $\vec{S}_t = \vec{S} \times \vec{S}_{xx}$ is obtained from
the $2\times 2$ matrix case by writing $\cS = \sum_{k=1}^3 S_k   \sigma_k$, where $\sigma_1, \sigma_2,\sigma_3$ are the Pauli matrices, and setting $t_2 = -\imag   t$.
\end{remark}

More generally, with the help of the Miura transformation (\ref{Miura_funct}) the
hierarchy equations resulting from (\ref{dualhier_funct2}) can be expressed solely in terms
of $\cS$. (\ref{Miura_funct}) implies
\begin{gather*}
    \la^{-1} \big\{ J , (g - g_{-[\la]})   g^{-1} \big\}
   =  \{ J , \cP \phi - \phi_{-[\la]} \cP \}
  = \left( \begin{array}{cc} 2 (p - p_{-[\la]}) & 0 \\ 0 & 0 \end{array} \right)
  = \left( \begin{array}{cc} - 2 \la   q   \bar{q}_{-[\la]} & 0 \\ 0 & 0 \end{array} \right) \\
\phantom{\la^{-1} \big\{ J , (g - g_{-[\la]})   g^{-1} \big\}}{}
   =  - 2 \la   \cP   [\cP , \phi ]   [\cP , \phi]_{-[\la]}  \cP
   = - 2 \la   \cP   g_x   g^{-1}   (g_x   g^{-1})_{-[\la]}  \cP    ,
\end{gather*}
where we assumed that (\ref{p_eq}) holds and used (\ref{dh_g-cP}) in the last step.
Multiplication from the left by $g^{-1}$ and from the right by $g$ leads to
\begin{gather}
    \la^{-1} \big\{ \cS , g^{-1} (g - g_{-[\la]}) \big\}
  = - \frac{1}{4} \la   (I+\cS)   \cS_x   g^{-1}   g_{x,-[\la]}
       g^{-1}_{-[\la]}   g   (I + \cS)   ,
           \label{dh_g-S_funct}
\end{gather}
where we used $g^{-1} \cP g = \frac{1}{2} (I + \cS)$, (\ref{dh_g^-1g_x->cS}),
and $\cS^2 =I$. At order $\la^1$ we obtain
\begin{gather*}
    \big\{ \cS , g^{-1} (g_{t_2} - g_{xx}) \big\}
  = - \frac{1}{2} (I+ \cS)   \cS_x   g^{-1} g_x   (I+\cS)
  = \frac{1}{4} (I+ \cS)   \cS_x{}^2   (I+\cS)
  = \frac{1}{2} \cS_x{}^2   (I+\cS)   ,
\end{gather*}
hence, using (\ref{dh_cS-ids}) and (\ref{dh_g^-1g_x->cS}),
\begin{gather}
    \big\{ \cS , g^{-1} g_{t_2} \big\}
   =  \big\{ \cS , g^{-1} g_{xx} \big\} + \frac{1}{2} \cS_x{}^2   (I+\cS)\nonumber\\
\phantom{\big\{ \cS , g^{-1} g_{t_2} \big\}}{}
   = \left\{ \cS , -\frac{1}{2} (\cS_x \cS)_x + \frac{1}{4} (\cS_x \cS)^2 \right\}
     + \frac{1}{2} \cS_x{}^2   (I+\cS)
   =  \frac{1}{2} \cS_x{}^2   .    \label{dh_cS_t2_anticomm}
\end{gather}
With the help of (\ref{dh_HMeq}), (\ref{cS_tn}) together with (\ref{dh_cS_t2_anticomm}) implies
\begin{gather}
     g^{-1} g_{t_2} = \frac{1}{2} \cS_{xx} + \frac{3}{4} \cS_x{}^2 \cS
                    =: \cF_2    .
                 \label{dh_g^-1g_t2->cS}
\end{gather}
At order $\la^2$, (\ref{dualhier_funct2}) yields
\begin{gather*}
    \cS_{t_3} = \left( \frac{3}{2} \cS_{t_2} - \frac{1}{2} \cS_{xx}
        + \frac{3}{2} \big(g^{-1} g_{t_2} + \big(g^{-1}\big)_{xx} g\big) (I+\cS) \right)_x   .
\end{gather*}
With the help of (\ref{dh_g^-1g_x->cS}) and (\ref{dh_g^-1g_t2->cS}), this can
be arranged into the form
\begin{gather*}
     \cS_{t_3}  =  \left( -\frac{3}{2} \cS_{xx} \cS
                 - \frac{3}{2} \cS_x{}^2
                 - \frac{1}{2} \cS_{xx}
                 + \frac{3}{2} \cS_{xx}   (I+\cS)
                 + \frac{3}{2} \cS_x{}^2   (I+\cS)
                 \right)_x \\
   \phantom{\cS_{t_3}}{}
          =  \left( \cS_{xx} + \frac{3}{2} \cS_x{}^2   \cS \right)_x
          = 2   \cF^{(2)}_x     .
\end{gather*}
In a similar way, from (\ref{dualhier_funct2}) and (\ref{dh_g-S_funct}) we obtain
\begin{gather*}
    g^{-1} g_{t_3}  =  -\frac{1}{2} (\cS_{xx} \cS)_x - \cS_x \cS_{xx}
         - \frac{5}{4} \cS_x{}^3 \cS =: \cF^{(3)}    , \\
    g^{-1} g_{t_4}  =  \frac{15}{8} \cS_x{}^2 \cS_{xx}
         + \frac{5}{4} \cS_x \cS_{xx} \cS_x
         + \frac{5}{4} \cS_x \cS_{xxx} \cS
         + \frac{5}{8} \cS_{xx} \cS_x{}^2
         + \frac{5}{4} \cS_{xx}{}^2 \cS   \\
    \phantom{g^{-1} g_{t_4}  =}{}
       + \frac{5}{4} \cS_{xxx} \cS_x \cS
         + \frac{35}{16} \cS_x{}^4 \cS
         + \frac{1}{2} \cS_{xxxx} =: \cF^{(4)}    ,
\end{gather*}
and
\begin{gather*}
     \cS_{t_n} = 2   \cF^{(n-1)}_x,  \qquad   n=2,3,4,5   ,
\end{gather*}
which likely extends to all higher $n \in \bbN$.
Of course, the expression for $\cS_{t_n}$ can be recovered by inserting
the corresponding expression for $g^{-1} g_{t_n}$ in~(\ref{cS_tn}).

\begin{remark}
Conditions like (\ref{J_anticom_g_xg^-1=0})\footnote{Via
a transformation $g \mapsto a   g$, with a suitably chosen invertible block-diagonal matrix $a$,
we can always achieve that (\ref{J_anticom_g_xg^-1=0}) holds (note that the
anticommutator of any matrix with $J$ is block-diagonal). Though this is a~symmetry of
the Heisenberg magnet equations (since it leaves $\cS$ invariant), it is not a~symmetry of the equations for~$g$. }
originated from the use of the Miura transformation, and they are in fact needed
to express the original hierarchy for the matrix variable $g$ in terms of $\cS$.
The ``mismatch'' in the Miura transformation, leading to the restriction
of the form of $g$, can be traced back to the fact that in the step from~(\ref{g_eq})
to~(\ref{Miura}) we are dropping cohomological terms (see Remark~\ref{rem:cohomology}).
\end{remark}

\begin{example}
If $m=2$, we write
\begin{gather*}
   g^{-1} = \left( \begin{array}{cc} \kappa & -\sigma \\ -\bar{\sigma} & \bar{\kappa} \end{array} \right)   ,  \qquad \mbox{hence} \qquad
    g = \frac{1}{\kappa \bar{\kappa} - \sigma \bar{\sigma}}
        \left( \begin{array}{cc} \bar{\kappa} & \sigma \\ \bar{\sigma} & \kappa \end{array} \right)   ,
\end{gather*}
and thus
\begin{gather*}
    \cS = g^{-1} J g = \frac{1}{\kappa \bar{\kappa} - \sigma \bar{\sigma}}
        \left( \begin{array}{cc} \kappa \bar{\kappa} + \sigma \bar{\sigma}
                     & 2 \kappa \sigma \\
                                 -2 \bar{\kappa} \bar{\sigma}
                     & -\kappa \bar{\kappa} - \sigma \bar{\sigma} \end{array} \right)   .
\end{gather*}
The condition (\ref{J_anticom_g_xg^-1=0}) amounts to
$\kappa_x   \bar{\kappa} - \sigma_x   \bar{\sigma} =0$ and
$\kappa   \bar{\kappa}_x - \sigma   \bar{\sigma}_x =0$. By adding these two equations,
we f\/ind that
\begin{gather}
     (\kappa \bar{\kappa} - \sigma \bar{\sigma})_x = 0   .  \label{m=2_kappa-sigma_conserv}
\end{gather}
\end{example}

\begin{remark}
Using the Miura transformation (\ref{Miura}), the linear equation (\ref{bidiff_linsys})
reads
\begin{gather*}
    \bd \Psi = \big[ (\bd g)   g^{-1} \big]   \Psi + 2 z   \d \Psi
\end{gather*}
(writing $\nu = 2 z$), hence
\begin{gather*}
    \bd \hat{\psi} = 2z   \big[\d \hat{\psi} + g^{-1} (\d g)   \hat{\psi} \big]
\end{gather*}
in terms of
\begin{gather*}
     \hat{\psi} = g^{-1}   \Psi   .
\end{gather*}
Evaluation of the linear equation for the bidif\/ferential calculus given by
(\ref{AKNS_bidiff_hier}) leads to
\begin{gather*}
   \la^{-1} \hat{\psi} - \la^{-1} \hat{\psi}_{-[\la]}   (I-2z   \la   \cP)
 = 2z   \big(g^{-1}_{-[\la]} \cP g\big)   \hat{\psi}   .
\end{gather*}
Writing
\begin{gather*}
     \hat{\psi} = \psi   e^{-\sum_{n \geq 1} (2z   \cP)^n   t_n}   ,
\end{gather*}
we obtain
\begin{gather*}
     \la^{-1} (\psi - \psi_{-[\la]}) = 2z   \big(g^{-1}_{-[\la]} \cP g\big)   \psi
   = z   \big(g^{-1}_{-[\la]}   g\big) (I+\cS)   \psi    .
\end{gather*}
Expansion in powers of $\la$ yields
\begin{gather*}
     \psi_x  =  z   (I+\cS)   \psi   , \\
     \psi_{t_2}  =  \big[ 2z^2   (I+\cS) - z   \cS_x \cS \big]   \psi   , \\
     \psi_{t_3}  =  \left[ 4z^3   (I+\cS) - 2z^2   \cS_x \cS
        + \frac{z}{2} \big(2 \cS_{xx} + 3 \cS_x{}^2 \cS\big) \right]   \psi   , \\
     \psi_{t_4}  =  \Bigg[ 8z^4   (I+\cS) - 4z^3   \cS_x \cS
        + z^2 \big( 2 \cS_{xx} + 3 \cS_x{}^2 \cS\big)\\
        \hphantom{\psi_{t_4}  =}{}
        - \frac{z}{2} \big(2 \cS_{xxx} \cS - 2 \cS_{xx} \cS_x - 4 \cS_x \cS_{xx}
               - 5 \cS_x{}^3 \cS\big) \Bigg]   \psi .
\end{gather*}
The f\/irst two equations constitute a Lax pair for the generalized Heisenberg magnet equation
(with $\cS^2 =I$).
\end{remark}

\subsection{A class of solutions}
\label{subsec:dh_solutions}
The following result is an analog to that in Section~\ref{subsec:NLShier_sol}
(see also Remark~3 in \cite{DMH08bidiff}). It allows to generate solutions of~(\ref{g_eq}) from solutions of a linear system.

\begin{theorem}
\label{theorem:dual}
Let $(\Omega, \d, \bar{\d})$ be a bidifferential calculus with
$\Omega = \cA \otimes \bigwedge(\bbC^N)$ and $\cA = \mathrm{Mat}_{n_0}(\cB)$,
for some $n_0 \in \bbN$. For f\/ixed $n \geq n_0$, let $\bS \in \mathrm{Mat}(n,n,\cB)$
and $\Delta \in \mathrm{Mat}(m,m,\cB)$. Furthermore, let $\bX \in \mathrm{Mat}(n,n,\cB)$
and $\bW \in \mathrm{Mat}(m,n,\cB)$ satisfy the linear equations
\begin{gather*}
   \bar{\d} \bX = (\d \bX)   \bS    , \qquad
   \bar{\d} \bW = (\d \bW)   \bS    ,
\end{gather*}
and also
\begin{gather}
    \bX   \bS - \bR   \bX = \tbV   \bZ   , \qquad
    \bW   \bS - \Delta   \bW = \bC   \bX    ,
            \label{bX,bW-algebraic-conds}
\end{gather}
with matrices $\bC, \bZ \in \mathrm{Mat}(m,n,\cB)$,
$\bR \in \mathrm{Mat}(n,n,\cB)$ and $\tbV \in \mathrm{Mat}(n,m,\cB)$, satisfying
\begin{gather*}
    \d \bC = 0   , \qquad
    \d \bR = 0   , \qquad
    \d \tbV = 0   , \qquad
    \bd \tbV = 0   .
\end{gather*}
Then
\begin{gather}
       g = \big(\bW   \bX^{-1} \tbV\big)^{-1}   ,    \label{g-solution}
\end{gather}
provided the inverse exists, solves the $($modified$)$ Miura transformation
equation \eqref{Miura}, i.e.
\begin{gather}
    [ \bd g - (\d g)   \Delta ]   g^{-1} = \d \phi   ,
    \label{mod_Miura}
\end{gather}
with some $m \times m$ matrix $\phi$, and thus $($by application of $\d)$ also \eqref{g_eq},
i.e.\footnote{This equation is invariant under right multiplication of~$g$
by any invertible $\d$- and $\bd$-constant matrix that commutes with $\Delta$.}
\begin{gather}
      \d \big( \left( \bd g - (\d g)   \Delta \right)   g^{-1} \big) = 0   .
           \label{mod_g_eq}
\end{gather}
\end{theorem}

\begin{proof} Using the Leibniz rule and the assumptions, we have
\begin{gather*}
  \bd g^{-1}  =  (\bd \bW)   \bX^{-1} \tbV - \bW   \bX^{-1} (\bd \bX)   \bX^{-1} \tbV \\
   \phantom{\bd g^{-1}}{}            =  (\d \bW)   \bS \bX^{-1} \tbV - \bW   \bX^{-1} (\d \bX)   \bS   \bX^{-1} \tbV
              =  \big( \d \bW - \bW   \bX^{-1} \d \bX \big)   \bS \bX^{-1} \tbV  \\
 \phantom{\bd g^{-1}}{}
              =  \d \big(\bW   \bX^{-1}\big)   \bX   \bS \bX^{-1} \tbV
              =  \d \big( \bW   \bS   \bX^{-1} \tbV \big)
                 - \bW   \bX^{-1}   \d \big( \bX   \bS \bX^{-1} \tbV \big) \\
 \phantom{\bd g^{-1}}{}
             =  \d \big(\Delta   g^{-1} \big) - \bW   \bX^{-1}   \d \big[ (\bR   \bX + \tbV \bZ)
                   \bX^{-1} \tbV \big]
              =  \d\big( \Delta   g^{-1}\big) - g^{-1}   \d \big(\bZ \bX^{-1} \tbV\big)   ,
\end{gather*}
and thus
\begin{gather*}
  [ \bd g - (\d g)   \Delta ]   g^{-1} = \d \big( \bZ \bX^{-1} \tbV - g   \Delta   g^{-1}\big)   .\tag*{\qed}
\end{gather*}\renewcommand{\qed}{}
\end{proof}

\begin{remark}
The assumptions in Theorem~\ref{theorem:dual} give rise to integrability conditions. The latter are
satisf\/ied if
\begin{gather*}
    \bd \bS = (\d \bS)   \bS   , \qquad
    \bd \Delta = (\d \Delta)   \Delta   , \qquad
    \bd \bC = (\d \Delta)   \bC   , \qquad
    \bd \bZ = (\d \bZ)   \bS   .
\end{gather*}
\end{remark}

In the following we exploit Theorem \ref{theorem:dual} for the bidif\/ferential calculus given
by~(\ref{AKNS_bidiff_hier}) with some simplif\/ications. We set $\Delta=0$ and make the further
assumption that $\bS$ is $\d$- and $\bd$-constant, and we write $\bZ = \tbU   \bY$, where
$\tbU \in \mathrm{Mat}(m,n,\cB)$ is $\d$- and $\bd$-constant and $\bY \in \mathrm{Mat}(n,n,\cB)$
solves $\bar{\d} \bY = (\d \bY)   \bS$. This is motivated by the fact that then the f\/irst
of conditions~(\ref{bX,bW-algebraic-conds}) reduces to~(\ref{Sylvester}), i.e.
\begin{gather}
     \bS   \bK - \bK   \bS = \bV   \bU   ,  \label{dh_Sylvester}
\end{gather}
assuming that $\bA_d$ is invertible and using results from Section~\ref{subsec:NLShier_sol},
in particular the def\/ini\-tions~(\ref{bK,bU,bV}) for $\bK$, $\bU$, $\bV$. Furthermore, we obtain
\begin{gather*}
     \bW = \bW_d + \bW_o   \bXi   , \qquad
     \bXi = e^{-\xi(\bS) \bJ}   , \qquad
     \xi(\bS) = \sum_{k \geq 1} \bS^k   t_k   ,
\end{gather*}
and the second of conditions~(\ref{bX,bW-algebraic-conds}) yields
$\bC = \bW_d \bS \bA_d^{-1}$ (which simply determines $\bC$)
and, assuming that $\bS$ is invertible,
\begin{gather*}
    \bW_o = - \bW_d   \bS   \bK   \bS^{-1}   .
\end{gather*}
The solution (\ref{g-solution}) of (\ref{mod_g_eq}) (with $\Delta=0$) is then given by
\begin{gather*}
     g^{-1} = \bW_d   \big(\bI - \bS   \bK   \bS^{-1} \bXi\big) (\bI - \bK   \bXi)^{-1} \bV   .
\end{gather*}
Rewriting this as
\begin{gather*}
    g^{-1}  =  \bW_d   \big(\bI - \bS   \bK   \bS^{-1} \bXi\big) (I+ \bK \bXi)
             \big(\bI - (\bK   \bXi)^2\big)^{-1} \bV  \\
\phantom{g^{-1}}{}
 =  \bW_d   \big(\bI - \bS   \bK   \bS^{-1} \bXi - \bS   \bK   \bS^{-1} \bXi \bK \bXi
              + \bK \bXi\big)\big(\bI - (\bK   \bXi)^2\big)^{-1} \bV   ,
\end{gather*}
it can easily be decomposed into a part that commutes with $J$,
\begin{gather*}
 \big(g^{-1}\big)_d  =  \bW_d   \big(\bI - \bS   \bK   \bS^{-1} \bXi \bK \bXi\big)
                   \big(\bI - (\bK   \bXi)^2\big)^{-1} \bV \\
 \phantom{\big(g^{-1}\big)_d}{}
             =  \bW_d   \bV + \bW_d   (\bK \bS - \bS   \bK)   \bS^{-1} \bXi \bK \bXi
                   \big(\bI - (\bK   \bXi)^2\big)^{-1} \bV \\
 \phantom{\big(g^{-1}\big)_d}{}
             =  \bW_d   \bV   \big[\bI - \bU   \bS^{-1} \bXi \bK \bXi
                   \big(\bI - (\bK   \bXi)^2\big)^{-1} \bV \big]   ,
\end{gather*}
and a part that anti-commutes with $J$,
\begin{gather*}
   \big(g^{-1}\big)_o  =  \bW_d   (\bK   \bS - \bS   \bK)   \bS^{-1}  \bXi
                    \big(\bI - (\bK   \bXi)^2\big)^{-1} \bV  \\
\phantom{\big(g^{-1}\big)_o }{}
               =  - \bW_d   \bV   \bU   \bS^{-1}  \bXi
                    \big(\bI - (\bK   \bXi)^2\big)^{-1} \bV   .
\end{gather*}
Using our concrete form of $J$ and $\bJ$, the matrices $\bK$, $\bS$, $\bU$, $\bV$
have the form given in (\ref{bK,bS,bU,bV_matrixform}), and we have
\begin{gather*}
    \bW_d = \left( \begin{array}{cc} W & 0 \\ 0 & \bar{W} \end{array} \right)   , \qquad
    \bXi =  \left( \begin{array}{cc} \Xi & 0 \\ 0 & \bar{\Xi} \end{array} \right)   , \qquad
    \Xi = e^{-\xi(S)}   , \qquad \bar{\Xi} = e^{\xi(-\bar{S})}   .
\end{gather*}
This leads to
\begin{gather*}
  g^{-1} = \left( \begin{array}{cc} \kappa & -\sigma \\ -\bar{\sigma} & \bar{\kappa} \end{array} \right)
\end{gather*}
where
\begin{gather*}
 \kappa = (W   V)   \big[I + U   \bar{S}^{-1} \bar{\Xi} \bar{K} \Xi
                (I-K \bar{\Xi} \bar{K} \Xi )^{-1} V \big]   , \\
 \bar{\kappa} = (\bar{W}   \bar{V})   \big[I + \bar{U}   S^{-1} \Xi K \bar{\Xi}
                (I- \bar{K} \Xi K \bar{\Xi} )^{-1} \bar{V} \big]   , \\
 \sigma = - (W   V)   U   \bar{S}^{-1} \bar{\Xi}   (I-\bar{K} \Xi K \bar{\Xi})^{-1} \bar{V}   , \\
 \bar{\sigma} = - (\bar{W}   \bar{V})   \bar{U}   S^{-1} \Xi
                (I- K \bar{\Xi} \bar{K} \Xi)^{-1} V   .
\end{gather*}
The only restrictions that have to be imposed on the matrices $K$, $\bar{K}$, $S$, $\bar{S}$, $U$, $\bar{U}$,
$V$, $\bar{V}$ result from (\ref{dh_Sylvester}). They are
\begin{gather}
    S K + K \bar{S} = VU   , \qquad
    \bar{S} \bar{K} + \bar{K} S = \bar{V} \bar{U}   .  \label{dh_Sylvester_decomp}
\end{gather}
The solutions of the hierarchy for $g$ obtained in this way also determine solutions of the
genera\-li\-zed Heisenberg hierarchy. This is so because the solutions constructed above via
Theorem~\ref{theorem:dual} are actually solutions of the Miura transformation and our choice
of matrix data via Proposition~\ref{prop:hier_sol} ensures that (\ref{p_eq}) holds
(which we used in Section~\ref{subsec:gHM}).

\subsection{Reciprocal dual and combined dual AKNS hierarchies}
Elaborating the dual equation (\ref{g_eq}) with the ``reciprocal'' bidif\/ferential calculus
determined by~(\ref{reciprocal_NLS_bidiff_hier}), instead of using that determined by~(\ref{AKNS_bidiff_hier}), we simply obtain~(\ref{dualhier_funct1}) with $g$ replaced
by~$g^{-1}$.
Again, we can combine the dual AKNS hierarchy and its reciprocal version, adopting
the procedure in Section~\ref{sec:comb_hier}. New equations arise from the mixed parts,
hence from evaluating~(\ref{g_eq}) using the bidif\/ferential calculus given by
\begin{gather*}
    \d f = [\cP \bbE_\la , f]   \zeta + \mu^{-1} [ \bar{\bbE}_\mu , f]   \bar{\zeta}   ,
           \qquad
    \bd f = \la^{-1} [\bbE_\la , f]   \zeta + [\cP \bar{\bbE}_\mu , f]   \bar{\zeta}   ,
\end{gather*}
which is a constituent of the calculus determined by (\ref{NLS_comb_bidiff}).
This results in
\begin{gather*}
    \big[\cP g_{-\bar{[\mu]}} \cP g^{-1} - \big(g_{-\bar{[\mu]}} \cP g^{-1}\big)_{-[\la]} \cP\big]
   - \mu^{-1} \la^{-1} \big[ g_{-[\la]} g^{-1} - \big( g_{-[\la]} g^{-1} \big)_{-\bar{[\mu]}} \big]
      = 0   .
\end{gather*}
To order $\la^0 \mu^0$ this is
\begin{gather*}
     \big(g_x g^{-1}\big)_{\bar{x}} + \big[\cP , g \cP g^{-1} \big] = 0   ,
\end{gather*}
hence
\begin{gather}
    \big(g_x g^{-1}\big)_{\bar{x}} = \frac{1}{4} \big( g \tilde{g}^{-1} - \tilde{g} g^{-1} \big)
    \qquad \mbox{where} \qquad
    \tilde{g} = J g J   .
             \label{gen_sG}
\end{gather}

The Miura transformation between the combined hierarchies consists of a pair of Miura
transformations, one for the original hierarchy and another one for the reciprocal.
It results in the following two generating equations,
\begin{gather*}
     \la^{-1} \big(g - g_{-[\la]}\big)   g^{-1} = \cP \phi - \phi_{-[\la]} \cP   , \qquad
     \big(\cP g - g_{-\bar{[\la]}} \cP\big)   g^{-1} = \la^{-1} \big(\phi - \phi_{-\bar{[\la]}}\big)   .
\end{gather*}
In particular, this yields
\begin{gather}
   g_x   g^{-1} = [\cP,\phi]   , \qquad
   \phi_{\bar{x}} = [\cP,g]   g^{-1}   .    \label{Miura_x,barx}
\end{gather}

The formulas in Section~\ref{subsec:dh_solutions} still generate solutions of
the combined dual hierarchy and also of the Miura transformation (cf.~(\ref{mod_Miura})),
provided we extend the expression for $\xi(S)$ used there~to
\begin{gather}
    \xi(S) = \sum_{k \geq 1} S^k   t_k + \sum_{k \geq 1} S^{-k}   \bar{t}_k   .
           \label{comb_dh_xi(S)}
\end{gather}

\subsubsection{Sine-Gordon solutions}
If $m=2$ and if $g$ has the form
\begin{gather*}
     g = f   \left( \begin{array}{cc} \cos(\vartheta/2) & - \sin(\vartheta/2) \\
           \sin(\vartheta/2) & \cos(\vartheta/2) \end{array} \right)
       = f   e^{\bbI   \vartheta/2}   , \qquad \bbI = \left( \begin{array}{cc} 0 & -1 \\
           1 & 0 \end{array} \right)   ,
\end{gather*}
where $f$ is a function independent of $x$,
then~(\ref{gen_sG}) becomes the sine-Gordon equation
\begin{gather*}
        \vartheta_{x \bar{x}} = \sin(\vartheta)   .
\end{gather*}
The function $f$ drops out of equation (\ref{gen_sG}). As a consequence of the form of $g$,
the condition~(\ref{J_anticom_g_xg^-1=0}), which arose from the Miura transformation,
is satisf\/ied. (\ref{Miura_x,barx}) requires the reduction conditions~(\ref{comb_hier_red1})
with $\epsilon =-1$ and then reads
\begin{gather*}
     q = - \frac{1}{2} \vartheta_x   , \qquad
     q_{\bar{x}} = - \frac{1}{2} \sin(\vartheta)   , \qquad
     p_{\bar{x}} = \frac{1}{2} [1 - \cos(\vartheta)]   .
\end{gather*}

In order to generate solutions of the sine-Gordon equation (and more generally of
the corresponding hierarchy), we have to choose
the matrix data in such a way that $g$ has the above form. We set
\begin{gather*}
     \bar{K} = - K     , \qquad
     \bar{S} = S       , \qquad
     \bar{U} = - U     , \qquad
     \bar{V} = V       , \qquad
     \bar{W} = W   .
\end{gather*}
Then (\ref{dh_Sylvester_decomp}) reduces to a single Sylvester equation, $S K + K S = VU$.
Setting $t_{2k} = \bar{t}_{2k} = 0$, $k=1,2,\ldots$, (\ref{comb_dh_xi(S)})
has the property $\xi(-S) = -\xi(S)$. As a consequence, we have
$\bar{\Xi} = e^{\xi(-S)} = e^{-\xi(S)} = \Xi$ and thus
\begin{gather*}
 \kappa = \bar{\kappa} = \alpha   \big[I - U S^{-1} \Xi K \Xi
                \big(I + (K \Xi)^2 \big)^{-1} V \big]  , \qquad
 \sigma = - \bar{\sigma} = - \alpha   U S^{-1} \Xi   \big( I + (K \Xi)^2 \big)^{-1} V   ,
\end{gather*}
where $\alpha = W   V$. We still have to ensure that $\kappa^2 + \sigma^2 = f^2$,
with some function $f$ that does not depend on $x$. But since our procedure actually
solves the Miura transformation (recall~(\ref{mod_Miura})), we already know that~(\ref{J_anticom_g_xg^-1=0}) is satisf\/ied,
hence~(\ref{m=2_kappa-sigma_conserv}) holds, which shows that $\kappa^2 + \sigma^2$ indeed
does not depend on $x$.

\begin{example}
Let $n=1$, $S = s \in \bbR$, and $U=V=W=1$. Then the Sylvester equation $S K + K S = VU$ is solved
by $K=\frac{1}{2s}$. Writing $\Xi = 2s   e^{-\tilde{\xi}(s)}$, where
$\tilde{\xi}(s) = \sum_{k \geq 0} (s^{2k+1} t_{2k+1} + s^{-2k-1} \bar{t}_{2k+1} ) + \xi_0$
with a constant $\xi_0$,
we obtain $\kappa = \tanh \tilde{\xi}(s)$ and $\sigma = \mathrm{sech} \tilde{\xi}(s)$,
so that $\kappa^2{+}\sigma^2{=}1$.
 From $\sigma/\kappa = \mathrm{csch} \tilde{\xi}(s)$ then follows the well-known kink solution
$\vartheta = 2   \arctan( \mathrm{csch} \tilde{\xi}(s)) = 4   \arctan e^{\tilde{\xi}(s)}$.
With $n>1$ and real diagonal $S$ we obtain multi-kink solutions.
\end{example}

\section{Conclusions}
\label{sec:concl}
We have shown in particular how a large family of solutions of matrix NLS equations, obtained
in \cite{DMH10NLS} with the help of general results of \cite{DMH08bidiff}, extends to solutions
of the corresponding hierarchies.

Moreover, by a simple exchange of the roles of $\d$ and $\bd$,
we obtained a ``reciprocal'' or ``purely negative'' counterpart of the AKNS hierarchy,
which turned out to be the nonlinear part of the potential KP hierarchy. Combining the two
hierarchies then gives rise to additional ``mixed f\/lows''. In this way we recovered in particular
the short pulse equation and obtained an apparently new vector version of it (dif\/ferent from those
considered in \cite{PKB08,Sakovich08}), for which we presented soliton solutions in the 2-component
case.

Via specialization of the general Miura transformation to the bidif\/ferential calculus studied in
this work, we recovered a relation between the AKNS hierarchy and the ``dual'' hierarchy of the generalized
Heisenberg magnet model. As the f\/irst ``mixed f\/low'' of the dual hierarchy combined with its
negative counterpart, with a certain reduction the sine-Gordon equation showed up.

In this work we concentrated on a simple method, introduced in \cite{DMH08bidiff}, to generate a class of
solutions, parametrized by certain matrix data (essentially of arbitrary size) subject to
a~Sylvester equation. The largest part of the work in \cite{DMH10NLS} concentrated on narrowing down
a~remaining redundancy in the matrix data that determine a matrix NLS solution. We expect
that most of these results can be carried over to the cases treated in the present work.

\subsection*{Acknowledgements}
We would like to thank Sergei Sakovich and some anonymous referees for helpful comments.

\pdfbookmark[1]{References}{ref}

\LastPageEnding

\end{document}